\documentclass[12pt,a4paper]{article}

\usepackage[backend=biber, style=numeric, citestyle=numeric-comp,
sorting=none, doi=true, eprint=true, url=true]{biblatex}
\usepackage{amsmath,amssymb,graphicx,subcaption,xcolor}

\usepackage{geometry}
\usepackage{comment}
\usepackage{enumitem}
\usepackage{caption}
\usepackage{tabularx}
\usepackage{booktabs}
\usepackage{makecell}
\usepackage{titlesec}

\usepackage{hyperref}
\hypersetup{colorlinks=true}

\usepackage{cleveref}
\usepackage{orcidlink}

\newcommand{\removed}[1]{}
\DeclareMathOperator{\sgn}{sgn}

\makeatletter
\newcommand\blfootnote[1]{%
	\begingroup
	\renewcommand\thefootnote{}\footnote{#1}%
	\addtocounter{footnote}{-1}%
	\endgroup
}
\makeatother

\titleformat{\paragraph}
{\normalfont\normalsize\bfseries}{\theparagraph}{1em}{}
\titlespacing*{\paragraph}
{0pt}{3.25ex plus 1ex minus .2ex}{1.5ex plus .2ex}

\titleformat{\section}
{\normalfont\Large\bfseries}{\thesection}{1em}{}

\titleformat{\subsection}
{\normalfont\large\bfseries}{\thesubsection}{1em}{}

\titleformat{\subsubsection}
{\normalfont\normalsize\bfseries}{\thesubsubsection}{1em}{}

\titlespacing*{\section}
{0pt}{3.5ex plus 1ex minus .2ex}{2.3ex plus .2ex}
\titlespacing*{\subsection}
{0pt}{3.25ex plus 1ex minus .2ex}{1.5ex plus .2ex}
\titlespacing*{\subsubsection}
{0pt}{3.25ex plus 1ex minus .2ex}{1.5ex plus .2ex}

\title{\bf Horizon, Null-Orbit, and Photon-Sphere Phase Structures in Logarithmic Nonlinear Electrodynamics}

\author{
	Peshwaz Abdulkareem Abdoul\,\orcidlink{0000-0002-2144-8336}
	\blfootnote{Email: peshwaz.abdoul@chu.edu.iq}
	\\
	{}\small Department of Physics, College of Science,
	Charmo University, \\
	\small Chamchamal, Sulaymaniyah, Kurdistan Region, Iraq
}

\date{}

\begin{document}
	
	\maketitle
	
	\begin{abstract}
		
	We investigate static, charged, asymptotically flat black holes in Einstein gravity minimally coupled to logarithmic nonlinear electrodynamics (LNED) in the $(Q,\epsilon)$ parameter space. We distinguish three related phase structures: the horizon geometry, the background circular null orbits, and the circular photon orbits of the effective optical geometry. All three share the same zero-radius non-critical boundary, $Q_{\rm nc}(\epsilon;M)=\gamma M^{2/3}\epsilon^{1/6}$, but have distinct critical structures: the horizon extremal curve $Q_{\rm ext}(\epsilon;M)$ and the background circular-null merger curve $Q_{\rm m}(\epsilon;M)$ both terminate at the bifurcation point $(Q_{\rm b},\epsilon_{\rm b})$, with $Q_{\rm b}=9M/[2\Gamma^2(1/4)]$ and $\epsilon_{\rm b}=128\pi^2Q_{\rm b}^2$, whereas the effective photon-orbit merger curve $Q_{\rm m}^{\rm ph}(\epsilon;M)$ does not pass through this point and has non-zero orbital radii for all $\epsilon$.
		
	For $\epsilon\ge\epsilon_{\rm b}$, approaching the $Q_{\rm nc}$ boundary shrinks the horizon to zero and produces black points with a non-zero quantum-corrected entropy $S_{\eta}(r_{\rm h}\to0)=\eta$. Their temperature vanishes at the bifurcation point and diverges for $\epsilon>\epsilon_{\rm b}$, except when $\eta\delta=1$, in which case it diverges at the bifurcation point as well. The background potential $V_{\rm eff}^{\rm bg}(r)$ develops an infinite barrier at $r=0$ without local extrema, acting as a perfect repeller. The photon potential $V_{\rm eff}^{\rm ph}(r)$ is regularized and keeps a maximum at finite radius, so it acts as a partial reflector. Thus the black points retain a photon sphere at finite radius, even though the background circular-null orbits have shrunk to zero.
		
	Away from $Q_{\rm nc}$, for $\epsilon<\epsilon_{\rm b}$ both the background and effective optical geometries support horizonless structures above $Q_{\rm ext}$, while for $\epsilon\ge\epsilon_{\rm b}$ only the effective optical geometry does so in the interval $Q_{\rm nc}<Q<Q_{\rm m}^{\rm ph}$. Both geometries develop an infinite barrier at the curvature singularity $r=0$, with a minimum and maximum at finite radius. Non-radial geodesics are blocked from the singularity and can form metastable bound states, while radial geodesics either reach $r=0$ or escape to infinity. The critical impact parameter encodes both the nonlinear gravitational and optical-geometry effects, providing an observational probe of the distinct phase structures.
		
	\end{abstract}
	

	\section{Introduction}
	
	Nonlinear electrodynamics (NED) extends Maxwell theory by allowing the electromagnetic Lagrangian to depend nonlinearly on the field invariant. Logarithmic nonlinear electrodynamics (LNED) is particularly useful because it admits an exact charged black-hole solution while recovering Maxwell electrodynamics in the weak-field limit. LNED black holes and related nonlinear charged geometries have been studied in connection with horizon structure, singularity behavior, thermodynamics, and particle dynamics~\cite{Soleng1995,Hendi2013,Kruglov2019,Luo2024,Brzo2026}. More generally, NED modifies not only the gravitational geometry but also the characteristic surfaces along which electromagnetic disturbances propagate, making the distinction between background null geodesics and physical photon trajectories essential for photon spheres and shadow-related observables. Related effects on perturbations and electromagnetic propagation have also been studied in charged NED spacetimes~\cite{Toshmatov2019,Breton2021}. The causality properties of such effective geometries have also been examined in detail for broad classes of NED models with a Maxwell limit~\cite{dePaula2024}.
	
	Null geodesics in charged black-hole spacetimes have been studied extensively in the linear Maxwell limit, both without rotation~\cite{Decanini2010} and in rotating cases~\cite{Chen2025}, providing a baseline for circular photon orbits in strong gravitational fields. Deviations from this baseline are precisely what nonlinear electrodynamics is expected to produce. Once nonlinear electrodynamics is introduced, high-frequency electromagnetic perturbations no longer follow the null geodesics of the background metric; they propagate instead on an effective optical metric built from derivatives of the nonlinear Lagrangian~\cite{Novello2000,DeLorenci2000,Novello2001}. Background circular-null orbits and physical photon spheres can therefore differ in radius and stability. Earlier work on exponential and logarithmic electrodynamics already revealed nontrivial photon and bound-orbit structure~\cite{Habibina2021}, and Tang et al.\ used the effective photon metric in slowly rotating logarithmic and exponential NED black holes to study shadow modifications~\cite{Tang2023}. Multiple photon spheres and stable inner photon configurations have also appeared in black-hole and NED settings~\cite{Guo2023}. Studies of circular orbits in regular and nonlinear-electrodynamic spacetimes show that strong-field corrections can qualitatively reshape both null and timelike trajectories~\cite{Stuchlik2015,Chiba2017,Toshmatov2026}. These results motivate a systematic comparison of the horizon, background circular-null, and effective photon-sphere structures within a single exact LNED solution. Systematic constructions of regular and horizonless spacetimes satisfying energy conditions and admitting photon spheres have also been developed in a complementary setting, where logarithmic and hypergeometric geometries play a central role and the Bardeen, Hayward, and Dymnikova models emerge as special cases~\cite{Battista2026,Wang2026}. The present LNED solution provides a contrasting case: it remains singular at $r	= 0$ yet still supports photon spheres and horizonless configurations, showing that regularity is not a prerequisite for the existence of a photon sphere or a critical impact parameter.
	
	The central issue is the relation between the phase boundaries of these three structures in the $(Q,\epsilon)$ parameter space. All three share the zero-radius non-critical boundary $C(\epsilon,Q;M)=0$, or equivalently $Q_{\rm nc}(\epsilon;M)=\gamma M^{2/3}\epsilon^{1/6}$, but their finite-radius critical curves are distinct. The horizon extremal curve $Q_{\rm ext}(\epsilon; M)$ and the background circular-null merger curve $Q_{\rm m}(\epsilon; M)$ both terminate at the same zero-radius bifurcation point $Q_{\rm b}=9M/[2\Gamma^2(1/4)]$, $\epsilon_{\rm b}=128\pi^2Q_{\rm b}^2$, whereas the effective photon-orbit merger curve $Q_{\rm m}^{\rm ph}(\epsilon;M)$ does not pass through this point. For $\epsilon<\epsilon_{\rm b}$ all three structures possess inner and outer branches: inner and outer horizons, inner stable and outer unstable background circular-null orbits, and inner stable and outer unstable effective photon orbits. On $Q=Q_{\rm nc}$ the inner branches reach the origin while the outer branches remain at finite radius; in addition, the background circular-null and effective photon-orbit structures both extend into the horizonless region, the latter reaching farther. For $\epsilon>\epsilon_{\rm b}$, the inner horizon and inner background circular-null branches disappear, and the background circular-null structure no longer extends into the horizonless region, whereas the effective photon-orbit structure retains both branches, so that for $Q_{\rm nc}<Q<Q_{\rm m}^{\rm ph}$ two effective photon orbits coexist in a horizonless configuration despite the absence of any finite positive background circular-null orbit. The disappearance of a background circular-null orbit therefore does not imply the disappearance of the physical photon structure in LNED. This behavior is consistent with the broader occurrence of circular-orbit structures in regular and horizonless compact spacetimes~\cite{Stuchlik2015,Chiba2017,Rayimbaev2020,Toshmatov2026} and with earlier LNED studies of stable photon configurations and multiple photon spheres~\cite{Habibina2021,Guo2023}.
	
	The effective potentials provide a complementary view of this distinction. On the non-critical boundary $Q=Q_{\rm nc}$, the background potential $V_{\rm eff}^{\rm bg}(r)$ develops an infinitely deep minimum for $\epsilon<\epsilon_{\rm b}$, whereas for $\epsilon\ge\epsilon_{\rm b}$ it develops an infinite barrier at $r=0$ and possesses no finite-radius extrema, acting as a perfect repeller. By contrast, the photon effective potential $V_{\rm eff}^{\rm ph}(r)$ is regularized and retains finite-radius extrema, acting only as a partial reflector. In the horizonless region between the non-critical and merger curves, both geometries develop an infinite barrier at the curvature singularity $r=0$, together with a minimum and a maximum at finite radius, corresponding to stable and unstable circular orbits, respectively. This barrier prevents non-radial geodesics from reaching the singularity, while radial null geodesics remain connected to infinity. The circular orbits at these extrema are metastable, and their instability timescales can be estimated through the Lyapunov exponent method, which connects the instability of circular null geodesics to the eikonal limit of quasinormal modes~\cite{Mondal2020,Wei2020}. Whether the metastable states found here support long-lived trapped modes, and whether they remain stable under gravitational and electromagnetic perturbations, requires a dedicated perturbation analysis that lies beyond the scope of this work.
	
	Causally, the infinite barrier at $r=0$ prevents non-radial geodesics from reaching the singularity, while radial null geodesics remain connected to infinity. This contrasts with regular black holes such as the Bardeen and Hayward solutions, where a de Sitter core replaces the singularity~\cite{AyonBeato1998,Hayward2006}, and it mirrors the extremal Reissner--Nordstr\"om case, whose horizon shrinks to zero and which is known to be unstable under linear perturbations~\cite{Aretakis2011}, with recent work providing further context for the dynamical fate of such zero-radius configurations~\cite{Angelopoulos2024}. Whether a different Lagrangian, quantum-gravity effects, or a dynamical collapse process could smooth out the singularity or form these horizonless configurations~\cite{Vertogradov2025, Battista2026b} remains an open question.
	
	These zero-radius configurations are black points whose thermodynamics can be analyzed directly. The bifurcation point itself, $(Q_{\rm b},\epsilon_{\rm b})$, is the Soleng-type black point, where $r_{\rm h}=0$ is a genuine horizon solution on $Q_{\rm nc}$; the ordinary black points, by contrast, are reached by approaching the boundary from the black-hole side, $Q<Q_{\rm nc}$ with $\epsilon>\epsilon_{\rm b}$, as $r_{\rm h}\to0$. Both carry a non-zero quantum-corrected entropy, $S_{\eta}(r_{\rm h}\to0)=\eta$, which replaces the vanishing uncorrected entropy $S_{0}\to0$ at the black point while the curvature singularity persists~\cite{Brzo2026}. Their temperature separates into two classes, in line with the general classification of black points in gravitating nonlinear electrodynamics~\cite{DiazAlonso2013}: it vanishes at the bifurcation point, where $f_{0}=0$, and diverges for $\epsilon>\epsilon_{\rm b}$, where $f_{0}\neq0$, except when the quantum-correction parameters satisfy $\eta\delta=1$, in which case the extremal temperature diverges logarithmically instead of vanishing. Because the black-point boundary ties the charge to the mass through $Q_{\rm nc}=\gamma M^{2/3}\epsilon^{1/6}$, the prefactor $f_{0}$ cannot be varied independently and approaches unity only in the small-mass limit, where the ordinary black points recover the Schwarzschild temperature, entropy, and heat capacity; the Soleng-type black points do not, since $f_{0}=0$ identically~\cite{Soleng1995}.
	
	The critical impact parameter translates this distinction into an observable. Introducing the optical factor $\Delta(r)=\sqrt{r^4+2\epsilon Q^2}$ that governs the effective angular sector, the impact parameter of the outer unstable effective photon sphere reads $b_{\rm c}^{\rm ph}=\sqrt{\Delta(r_{\rm ph})/f(r_{\rm ph})}$, whereas for the background circular-null orbit, $b_{\rm c}^{\rm bg}=r_{\rm c}/\sqrt{f(r_{\rm c})}$. This difference reflects both the displacement of the physical photon-sphere radius relative to the background circular-null radius and the modification of the angular sector by the effective optical geometry, and it complements previous shadow calculations in logarithmic NED, including the slowly rotating configurations of Tang et al.~\cite{Tang2023}. Similar enhancements of the critical impact parameter in the effective geometry, relative to the background geometry, have been reported for other electrically charged NED regular black holes~\cite{dePaula2023}, and the relation between effective-geometry photon motion, gravitational redshift, and observable signatures has been explored in~\cite{dePaula2025}. For black-hole configurations, the outer unstable photon sphere and its critical impact parameter determine the characteristic scale for photon capture and shadow formation under appropriate observer conditions. For the black points, the effective photon sphere survives at finite radius even though the corresponding background circular orbit has collapsed to $r=0$, so that a finite critical impact parameter — and hence a finite shadow — persists in a regime where a background analysis would predict its disappearance. For a large part of our horizonless sectors, an outer photon sphere and an inner stable photon orbit coexist, so that a finite shadow is accompanied by a region of stable circular photon motion — except on the $Q_{\rm nc}$ boundary for $\epsilon > \epsilon_{\rm b}$, where the inner orbit shrinks to zero. Whether the interior is regular, as in the Bardeen and Hayward solutions, or singular, as in the present LNED case, does not affect the existence of the shadow; it affects only what the shadow conceals. Horizonless regular spacetimes and NED effective geometries have indeed been shown to support photon spheres and circular photon orbits even without an event horizon~\cite{Stuchlik2015,Toshmatov2026}.
	
	This paper is organized as follows. In Sec.~\ref{sec:action_integral}, we introduce the Einstein--LNED action and derive the coupled field equations. Sec.~\ref{sec:field_equations} presents the exact static, spherically symmetric charged solution and its Maxwell and near-origin limits. In Sec.~\ref{sec:horizon_phase}, we analyze the horizon phase structure, derive the extremal and non-extremal boundaries together with their bifurcation point, and classify the resulting black-point configurations together with their thermodynamic properties. Sec.~\ref{sec:PMBG} treats the background circular null geodesics, their merger structure, and the associated effective potential. Sec.~\ref{sec:PMEG} constructs the effective optical geometry, analyzes the physical photon spheres and the photon effective potential, and compares the critical impact parameters of the two geometries. Finally, we summarize the implications of the distinct gravitational and optical nonlinearities for photon propagation.

	\section{Action Integral and Field Equations}\label{sec:action_integral}
	
	In this section, we consider Einstein gravity coupled to logarithmic nonlinear electrodynamics (LNED), for which the action is given by
	\begin{equation}\label{eq:action}
		S = \int d^4x \sqrt{-g} \left[ \frac{1}{2\kappa}(R - 2\Lambda) - \mathcal{L}(\mathcal{F}) \right],
	\end{equation}
	where the Lagrangian density of nonlinear electrodynamics, $\mathcal{L}(\mathcal{F})$, is given by~\cite{Brzo2026}
	\begin{equation}\label{eq:L_electro}
		\mathcal{L}(\mathcal{F}) = - \frac{1}{\epsilon}\ln\bigg(1 + \frac{\epsilon}{4} \mathcal{F}\bigg),
	\end{equation}
	where $\epsilon$ is the nonlinear coupling parameter, $\Lambda$ is the cosmological constant, and $\kappa = 8\pi$\footnote{Throughout this work, we adopt Planck units, with $G=c=k_{B}=4\pi\epsilon_{0}=\hbar=1$.}. The electromagnetic invariant is defined as $\mathcal{F} = F_{\mu\nu}F^{\mu\nu}$, where
	\begin{equation}\label{eq:field_strength}
		F_{\mu\nu} = \partial_{\mu} A_{\nu} - \partial_{\nu} A_{\mu}
	\end{equation}
	is the electromagnetic field-strength tensor. The Lagrangian~\eqref{eq:L_electro} is well defined provided $1+\epsilon\mathcal{F}/4>0$, which, for a purely electric configuration, translates into a bound on the field strength that will be made explicit in Sec.~\ref{sec:field_equations}. In the limit $\epsilon \to 0$, the Lagrangian density in Eq.~\eqref{eq:L_electro} reduces to the standard Maxwell form,
	\begin{equation}\label{eq:maxwell_limit}
		\mathcal{L}(\mathcal{F}) \approx - \frac{1}{4}\mathcal{F}
		= -\frac{1}{4}F_{\mu\nu}F^{\mu\nu}.
	\end{equation}
	We use this convention throughout, so that the effective-metric formula employed in Sec.~\ref{sec:PMEG} matches that of Refs.~\cite{Novello2000,DeLorenci2000}.
	
	Varying the action in Eq.~\eqref{eq:action} with respect to the metric $g_{\mu\nu}$ yields the Einstein field equations,
	\begin{equation}\label{eq:einstein_eq}
		G_{\mu\nu} + \Lambda g_{\mu\nu} = R_{\mu\nu} - \frac{1}{2} g_{\mu\nu} R + \Lambda g_{\mu\nu} = \kappa T_{\mu\nu},
	\end{equation}
	where $G_{\mu\nu}$ is the Einstein tensor and $R$ is the Ricci scalar, defined as the trace of the Ricci tensor $R_{\mu\nu}$, i.e.,
	$R = g^{\mu\nu}R_{\mu\nu}$. The stress-energy tensor associated with the nonlinear electromagnetic field is
	\begin{equation}\label{eq:stress_energy}
		T_{\mu\nu} = \frac{F_{\mu}{}^{\alpha} F_{\nu\alpha}}{1+\frac{\epsilon}{4} \mathcal{F}} - \frac{g_{\mu\nu}}{\epsilon}\ln\bigg(1+\frac{\epsilon}{4} \mathcal{F}\bigg),
	\end{equation}
	which follows from $T_{\mu\nu} = -4\mathcal{L}_{\mathcal{F}}F_{\mu}{}^{\alpha}F_{\nu\alpha} + g_{\mu\nu}\mathcal{L}$, with $\mathcal{L}_{\mathcal{F}} = \partial\mathcal{L}/\partial\mathcal{F}$.
	
	Furthermore, variation of the action with respect to the electromagnetic four-potential $A_{\mu}$ yields the nonlinear electromagnetic field equations,
	\begin{equation}\label{eq:electro_eq}
		\nabla_\mu \left( \frac{F^{\mu\nu}}{1+\frac{\epsilon}{4} \mathcal{F}} \right) = 0,
	\end{equation}
	which follow from the Euler--Lagrange equation $\nabla_\mu(\mathcal{L}_{\mathcal{F}}F^{\mu\nu})=0$. In the limit $\epsilon \to 0$, the stress-energy tensor and the electromagnetic field equations reduce to their respective Maxwell forms,
	\begin{equation}\label{eq:maxwell_forms}
		T_{\mu\nu} \approx F_{\mu}{}^{\alpha}F_{\nu\alpha}
		-\frac{1}{4}g_{\mu\nu}F_{\alpha\beta}F^{\alpha\beta},
		\qquad
		\nabla_\mu F^{\mu\nu} \approx 0.
	\end{equation}
	In the following section, we obtain exact solutions to the coupled field equations~\eqref{eq:einstein_eq} and~\eqref{eq:electro_eq} for a static, spherically symmetric, purely electric configuration.

	\section{Exact Solutions to the Field Equations}\label{sec:field_equations}
	
	In this work, we restrict our analysis to static, spherically symmetric black holes with a purely electric electromagnetic field. The spacetime line element is given by
	\begin{equation}\label{eq:metric}
		ds^2 = -f(r) dt^2 + \frac{dr^2}{f(r)} + r^2(d\theta^2 + \sin^2\theta d\varphi^2),
	\end{equation}
	where the metric tensor $g_{\mu\nu}$ is diagonal, with nonvanishing components
	\begin{equation}\label{eq:metric_components}
		g_{tt} = -f(r), \qquad
		g_{rr} = \frac{1}{f(r)}, \qquad
		g_{\theta\theta}=r^{2}, \qquad
		g_{\varphi\varphi}=r^{2}\sin^{2}\theta.
	\end{equation}
	Neglecting the magnetic field, we take the electromagnetic four-potential to be
	\begin{equation}\label{eq:four_potential}
		A_{\mu} = \big(\Phi(r),0,0,0\big).
	\end{equation}
	Consequently, the only nonvanishing components of the electromagnetic field-strength tensor are
	\begin{equation}\label{eq:field_strength_components}
		F_{tr}=-F_{rt}=E(r),
	\end{equation}
	where all other components vanish. The electromagnetic invariant is therefore
	\begin{equation}\label{eq:invariant}
		\mathcal{F}
		=F_{\mu\nu}F^{\mu\nu}
		=2F_{tr}F^{tr}
		=2F_{tr}\left(g^{tt}g^{rr}F_{tr}\right)
		=-2\big(E(r)\big)^2.
	\end{equation}
	Here, $E(r)$ and $\Phi(r)$ denote the radial electric field and electric potential, respectively, and they are related by
	\begin{equation}\label{eq:electric_field_potential}
		E(r)=-\frac{d\Phi(r)}{dr}.
	\end{equation}
	
	Using the metric~\eqref{eq:metric}, the determinant of the metric tensor is $g=-r^4\sin^2\theta$, and hence $\sqrt{-g}=r^2\sin\theta$. The nonlinear electromagnetic field equations~\eqref{eq:electro_eq} can be written in the form
	\begin{equation}\label{eq:electro_eq_expanded}
		\frac{1}{\sqrt{-g}}\partial_\mu
		\left[
		\sqrt{-g}
		\frac{F^{\mu\nu}}
		{1+\frac{\epsilon}{4}\mathcal{F}}
		\right]=0.
	\end{equation}
	For the purely electric configuration considered here, the only nontrivial component is the $\nu=t$ component. Using $F^{rt}=E(r)$ and $\mathcal{F}=-2E^2$, we obtain
	\begin{equation}\label{eq:reduced_electro_eq}
		\partial_r \left(
		\frac{r^2 E}{1-\frac{\epsilon}{2} E^{2}}
		\right) = 0.
	\end{equation}
	The factor $r^2$ originates from the determinant of the spherically symmetric metric, or equivalently from the area factor of the spherical surfaces of constant radius. Integrating Eq.~\eqref{eq:reduced_electro_eq} gives
	\begin{equation}\label{eq:charge_cons}
		\frac{r^2 E}{1-\frac{\epsilon}{2} E^{2}} = Q,
	\end{equation}
	where $Q$ is an integration constant associated with the electric charge. Solving the resulting quadratic equation for $E(r)$ yields
	\begin{equation}\label{eq:electric_field}
		E(r) = \frac{\sqrt{r^{4} + 2\epsilon Q^2} - r^{2}}{\epsilon Q}.
	\end{equation}
	The branch displayed above is selected because it continuously recovers the Coulomb field in the weak-field (asymptotic) limit. In particular, for $r\to\infty$,
	\begin{align}\label{eq:electric_field_asymptotic}
		E(r) \approx \frac{Q}{r^{2}} - \frac{\epsilon Q^{3}}{r^{6}} + \mathcal{O}\left(\frac{\epsilon^2 Q^5}{r^{10}}\right).
	\end{align}
	Thus, the leading-order term is precisely the Coulomb electric field. On the other hand, Eq.~\eqref{eq:electric_field} gives the following behavior near the origin:
	\begin{equation}\label{eq:electric_field_origin}
		E(r) \approx \sqrt{\frac{2}{\epsilon}}\,\sgn(Q)
		- \frac{r^{2}}{\epsilon Q} + \mathcal{O}(r^4).
	\end{equation}
	Therefore, for $\epsilon>0$, the electric field remains finite at the origin, with
	\begin{equation}\label{eq:electric_field_limit}
		\lim_{r\to0}E(r)=\sqrt{\frac{2}{\epsilon}}\,\sgn(Q).
	\end{equation}
	This behavior demonstrates that logarithmic nonlinear electrodynamics removes the $1/r^2$ divergence of the electric field present in Maxwell electrodynamics in the limit $\epsilon\to0$~\cite{Soleng1995, Kruglov2019}. Integrating Eq.~\eqref{eq:electric_field}, the electric potential,
	\begin{equation}\label{eq:electric_potential_integral}
		\Phi(r) = - \int^{r} E(r^{\prime})\,dr^{\prime},
	\end{equation}
	can, after some algebraic manipulation, be written as
	\begin{equation}\label{eq:electric_potential}
		\Phi(r) = \frac{r^{3} - r\sqrt{r^{4} + 2\epsilon Q^2}}{3\epsilon Q}
		-\frac{4Q}{3} \int^{r} \frac{dr^{\prime}}{\sqrt{r^{\prime 4}+2\epsilon Q^{2}}}
		+\Phi(\epsilon,Q),
	\end{equation}
	where $\Phi(\epsilon,Q)$ is an integration constant independent of $r$. The integral on the right-hand side can be expressed in terms of the Gaussian hypergeometric function as
	\begin{equation}\label{eq:hypergeometric_integral}
		\int^{r} \frac{dr^{\prime}}{\sqrt{r^{\prime 4}+2\epsilon Q^{2}}}
		= \frac{r}{\sqrt{2\epsilon}|Q|}
		{}_2F_1\left(\tfrac{1}{4},\tfrac{1}{2};\tfrac{5}{4};
		-\frac{r^4}{2\epsilon Q^2}\right).
	\end{equation}
	Consequently, the exact electric potential is
	\begin{equation}\label{eq:electric_potential_exact}
		\Phi(r) =
		\frac{r^{3} - r\sqrt{r^{4} + 2\epsilon Q^2}}{3\epsilon Q}
		-\sqrt{\frac{8}{\epsilon}}\frac{\sgn(Q)}{3}\,
		r\left[
		{}_2F_1\left(\tfrac{1}{4},\tfrac{1}{2};\tfrac{5}{4};
		-\frac{r^4}{2\epsilon Q^2}\right)
		\right]
		+\Phi(\epsilon,Q).
	\end{equation}
	
	It is important to emphasize that $\Phi(\epsilon,Q)$ is an integration constant and is independent of $r$. It represents the freedom to choose the zero of the electric potential. If we impose the gauge condition $\Phi(r)\to0$ as $r\to\infty$, this constant is fixed to be
	\begin{equation}\label{eq:phi_constant}
		\Phi(\epsilon,Q)
		= \Phi_{0}
		+\frac{[\Gamma(1/4)]^{2}}{3\sqrt{\pi}\,2^{1/4}}
		\frac{\sgn(Q)\sqrt{|Q|}}{\epsilon^{1/4}},
	\end{equation}
	where $\Phi_{0}$ is an arbitrary constant associated with the residual gauge freedom. In the gauge $\Phi_{0}=0$, one has $\Phi(\infty)=0$.
	
	We now turn to the determination of the metric function $f(r)$. It follows from the $tt$ component of the Einstein equations, $G_{tt}+\Lambda g_{tt}=\kappa T_{tt}$, that, for the metric ansatz~\eqref{eq:metric} and the purely electric field considered above,
	\begin{align}\label{eq:Gtt}
		G_{tt} -\Lambda f
		= \frac{f}{r^{2}}
		\left(1-f-r\frac{df}{dr}-\Lambda r^{2}\right).
	\end{align}
	Furthermore, using Eq.~\eqref{eq:charge_cons} to replace
	$1-\frac{\epsilon}{2}E^{2}$ by $\frac{r^{2}E}{Q}$, Eq.~\eqref{eq:stress_energy} gives
	\begin{align}\label{eq:Ttt}
		T_{tt}
		= f \frac{QE}{r^{2}}
		+ f \left[\frac{1}{\epsilon}
		\ln\left(\frac{r^{2}E}{Q}\right)\right].
	\end{align}
	Therefore, the Einstein equation yields the following first-order differential equation for the metric function $f(r)$:
	\begin{align}\label{eq:diff_f}
		r\frac{df}{dr} + f
		= 1 - \Lambda r^{2} - \kappa r^{2}\rho(r),
	\end{align}
	where $\rho(r)$ is the effective energy density, defined by
	\begin{align}\label{eq:rho}
		\rho(r) = -T_{t}^{t}
		= -g^{tt}T_{tt}
		= \frac{T_{tt}}{f}
		= \frac{QE}{r^{2}}
		+\frac{1}{\epsilon}
		\ln\left(\frac{r^{2}E}{Q}\right).
	\end{align}
	Solving Eq.~\eqref{eq:diff_f} using the integrating-factor method gives
	\begin{align}\label{eq:f_r}
		f(r)
		= 1 - \frac{2C(\epsilon,Q;M)}{r}
		-\frac{\Lambda r^{2}}{3}
		-\frac{\kappa}{r}\int^{r}r^{\prime 2}\rho(r^{\prime})\,dr^{\prime}.
	\end{align}
	Although the integral on the right-hand side is lengthy, it can be evaluated exactly. The resulting metric function is
	\begin{align}\label{eq:exact_fr}
		f(r) &= 1 - \frac{2C(\epsilon,Q;M)}{r} -\frac{\Lambda  r^{2}}{3}
		- \frac{5\kappa }{9 \epsilon}
		\bigg(\sqrt{r^{4} + 2\epsilon Q^2} - r^2\bigg)
		- \frac{\kappa r^{2}}{3 \epsilon}
		\ln\bigg(
		\frac{r^2\sqrt{r^4 + 2\epsilon Q^2} - r^{4}}
		{\epsilon Q^2}
		\bigg)
		\nonumber \\
		&\quad
		- \sqrt{\frac{2}{\epsilon}}\frac{4|Q|\kappa}{9}
		\left[
		{}_2F_1\left(\tfrac{1}{4},\tfrac{1}{2};\tfrac{5}{4};
		-\frac{r^4}{2\epsilon Q^2}\right)
		\right].
	\end{align}
	The integration constant $C(\epsilon,Q;M)$ is related to the physical mass of the black hole. To determine its normalization, we require that, for $\Lambda=0$, the metric function approaches the Reissner--Nordstr\"om (RN) solution at large distances. Introducing the conventional charge parameter $q$ through
	\begin{equation}\label{eq:charge_parameter}
		4\pi Q^2=q^2,
	\end{equation}
	the asymptotic behavior is
	\begin{align}\label{eq:RN_asymptotic}
		f(r) \approx 1 - \frac{2M}{r} + \frac{q^{2}}{r^{2}}
		+\mathcal{O}\left(\frac{\epsilon}{r^{6}}\right).
	\end{align}
	This asymptotic Reissner--Nordstr\"om form is the same weak-field limit that characterizes the Hendi-type logarithmic black-hole family~\cite{Hendi2013}, to which the present solution belongs. It fixes the mass parameter
	$M$ that appears in the integration constant below.
	\begin{align}\label{eq:norm_mass}
		C(\epsilon,Q;M)
		= M
		-\frac{2^{3/4}\kappa [\Gamma(1/4)]^{2}}
		{18\sqrt{\pi}}
		\frac{|Q|^{3/2}}{\epsilon^{1/4}}.
	\end{align}
	Here, $M$ denotes the physical mass parameter of the black hole, whereas $C(\epsilon,Q;M)$ is an integration constant that depends on $\epsilon$ and $Q$ but remains constant with respect to $r$. The difference between $M$ and $C$ accounts for the finite electromagnetic self-energy associated with the nonlinear electric field, which scales as $\epsilon^{-1/4}|Q|^{3/2}$ and remains finite for every nonzero $\epsilon$; in this sense, nonlinear electrodynamics regularizes a self-energy that would otherwise diverge in the Maxwell theory. The curvature invariants, however, remain singular at $r=0$, as in the Maxwell case~\cite{Soleng1995}.

	\section{Horizon Phase Structure and Bifurcation}\label{sec:horizon_phase}
	
	The horizons of the spacetime are determined by the positive real roots of
	\begin{equation}\label{eq:horizon_condition}
		f(r_{\rm h})=0.
	\end{equation}
	Depending on the parameters $(M,Q,\epsilon,\Lambda)$, the spacetime can exhibit different horizon structures. These include a black hole with a single non-degenerate outer horizon, a black hole with two distinct horizons corresponding to an inner and an outer horizon, and a horizonless configuration. An analytical solution of $f(r_{\rm h})=0$ for the horizon radius is generally difficult to obtain because of the nonlinear and transcendental form of the metric function. Nevertheless, important features of the horizon phase structure can be obtained analytically from the extremality condition. A degenerate horizon satisfies
	\begin{equation}\label{eq:extremality_conditions}
		f(r_{\rm h}^{\rm ext})=0,
		\qquad
		\frac{df(r)}{dr}
		\Big.\Big|_{r = r_{h}^{\rm ext}}=0.
	\end{equation}
	These conditions can be applied directly to Eq.~\eqref{eq:diff_f}. Evaluating that equation at $r=r_{\rm h}^{\rm ext}$ and imposing both extremality conditions yields
	\begin{align}\label{eq:ext_BH}
		H(r_{\rm h}^{\rm ext},Q,\epsilon,\Lambda)
		=
		1-\Lambda \left[r_{\rm h}^{\rm ext}\right]^{2}
		-\kappa \left[r_{\rm h}^{\rm ext}\right]^{2}
		\rho(r = r_{\rm h}^{\rm ext},Q,\epsilon)
		=0.
	\end{align}
	where $\rho(r = r_{\rm h}^{\rm ext})$ is given by Eq.~\eqref{eq:rho}. This defines the extremal curve $Q_{\rm ext}(\epsilon;M)$ in the $(Q,\epsilon)$ plane. In the Maxwell limit $\epsilon\to0$, the energy density becomes
	\begin{equation}\label{eq:rho_maxwell}
		\rho(r_{\rm h}^{\rm ext})
		\sim
		\frac{Q^2}
		{2	\left[r_{\rm h}^{\rm ext}\right]^{4}}.
	\end{equation}
	Using $q^2=4\pi Q^2$, Eq.~\eqref{eq:ext_BH} reduces to the familiar Reissner--Nordstr\"om (RN) extremality condition
	\begin{align}\label{eq:ext_BH_RN}
		H_{0}(r_{\rm h}^{\rm ext},q,\Lambda)
		=1-\Lambda \left[r_{\rm h}^{\rm ext}\right]^{2} -\frac{q^{2}}{\left[r_{\rm h}^{\rm ext}\right]^{2}}
		=0.
	\end{align}
	For $\Lambda=0$, this gives $r_{\rm h}^{\rm ext}=q$, showing that the RN extremal configuration occurs at a finite horizon radius and corresponds to the merger of the inner and outer horizons. This finite-radius extremal branch is a general feature of charged NED black holes, including dyonic and magnetically charged configurations~\cite{Kruglov2019,Luo2024}.
	\begin{figure}[t]
		\centering
		\includegraphics[width=0.7\textwidth]{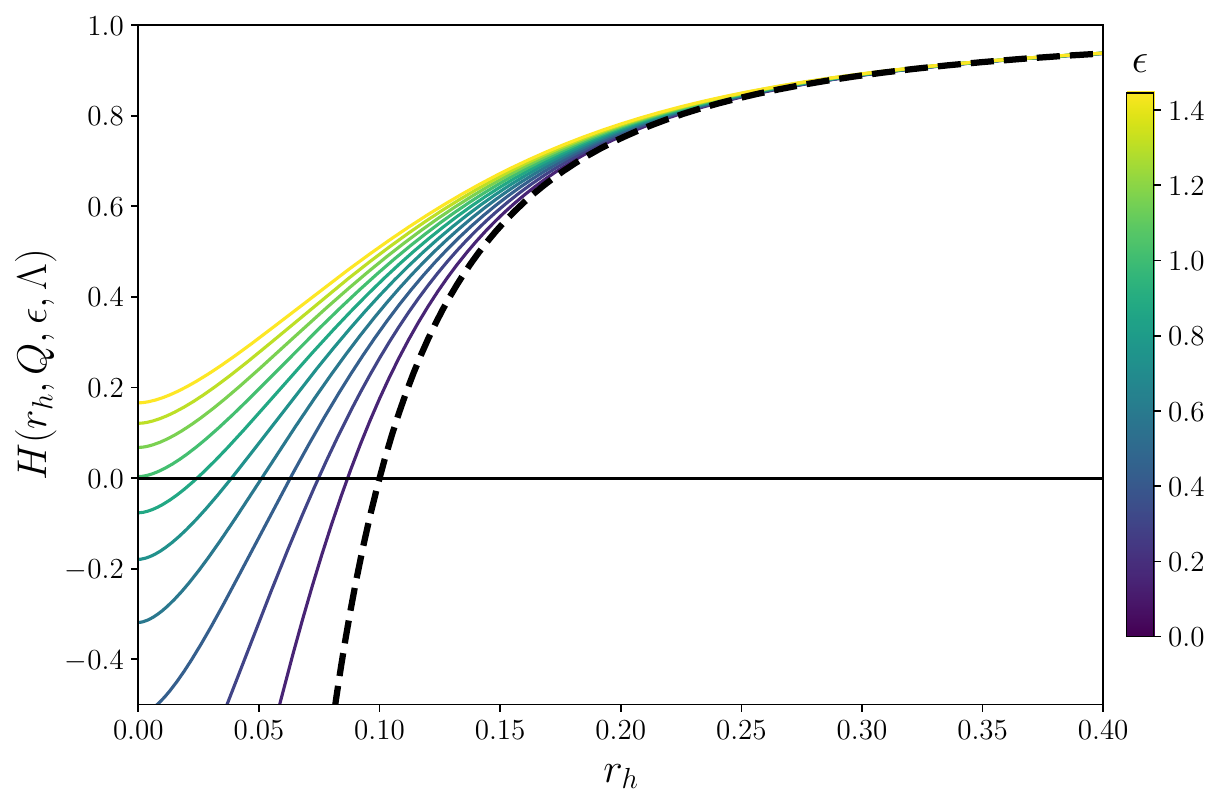}
		\caption{The extremal function $H(r_{\rm h},\epsilon,Q,\Lambda)$ from Eq.~\eqref{eq:ext_BH} is plotted against the horizon radius $r_{\rm h}$ for different values of $\epsilon$, with extremal charge $q=\sqrt{4\pi}\,Q=0.1$ and $\Lambda=0$. The Reissner--Nordstr\"om solution $H_0(r_{\rm h},q,\Lambda)$ is represented by the dashed curve, corresponding to $\epsilon=0$. The intersections of the curves with the $r_{\rm h}$-axis, i.e., the positive roots of $H(r_{\rm h})=0$, determine the corresponding extremal horizon radii $r_{\rm h}^{\rm ext}$.}
		\label{fig:extremal}
	\end{figure}

	The nonlinear theory exhibits qualitatively different behavior as the nonlinear coupling parameter $\epsilon$ increases. As shown in Fig.~\ref{fig:extremal}, the extremal function $H(r_{\rm h},\epsilon,Q,\Lambda)$ develops an $\epsilon$-dependent positive root that determines the extremal horizon radius through $H(r_{\rm h}^{\rm ext})=0$. In the Maxwell limit, $\epsilon\rightarrow0$, the corresponding root reproduces the RN extremal radius. As $\epsilon$ increases, this positive root shifts toward smaller values of $r_{\rm h}$, indicating that the extremal horizon becomes progressively smaller. At a critical value of the nonlinear coupling, the positive root reaches the origin, and the finite-radius extremal branch terminates. Thus, the endpoint of the extremal branch is characterized by
	\begin{equation}\label{eq:extremal_radius_limit}
		r_{\rm h}^{\rm ext}\rightarrow0.
	\end{equation}
	The disappearance of the positive root of $H$ at this point marks the transition between the finite-radius extremal regime and the zero-radius bifurcation point. To determine the corresponding bifurcation parameters, we evaluate Eq.~\eqref{eq:ext_BH} in the limit $r_{\rm h}^{\rm ext}\to0$. Knowing that 
	\begin{equation}
		\lim_{r_{h}^{\rm ext} \to 0} \rho(r_{h}^{\rm ext}) 
		= 
		\frac{|Q_{b}|}{\left[r_{h}^{\rm ext}\right]^{2}} \sqrt{\frac{2}{\epsilon_{b}}}, 
	\end{equation}
	for $\Lambda=0$ and $\kappa = 8\pi$, this yields
	\begin{align}\label{eq:eps_cr}
		\epsilon_{\rm b}=128\pi^2 Q_{\rm b}^2,
	\end{align}
	where $Q_{\rm b}$ and $\epsilon_{\rm b}$ denote the charge and nonlinear coupling at the bifurcation point, respectively. This relation shows that the critical nonlinear coupling is directly tied to the charge at which the finite-radius extremal branch terminates. At the bifurcation point, the horizon radius vanishes and the horizon condition in Eq.~\eqref{eq:exact_fr} reduces to
	\begin{equation}\label{eq:C_zero}
		C(\epsilon_{\rm b},Q_{\rm b};M)=0.
	\end{equation}
	Combining this condition with Eq.~\eqref{eq:norm_mass}, and using Eq.~\eqref{eq:eps_cr}, yields the bifurcation charge in terms of the physical mass,
	\begin{align}\label{eq:QM_cr}
		Q_{\rm b}(M)
		=
		\frac{9M}{2[\Gamma(1/4)]^2}.
	\end{align}
	Substitution of Eq.~\eqref{eq:QM_cr} into Eq.~\eqref{eq:eps_cr} gives the corresponding bifurcation value of the nonlinear coupling,
	\begin{align}\label{eq:eps_M_cr}
		\epsilon_{\rm b}(M)
		= \frac{6^{5} \pi^2 M^2}{3[\Gamma(1/4)]^4}.
	\end{align}
	Thus, the bifurcation parameters scale as $Q_{\rm b}\propto M$ and $\epsilon_{\rm b}\propto M^2$. Consequently, decreasing the black-hole mass shifts the bifurcation point toward smaller values of both $Q$ and $\epsilon$.	We next determine the second boundary of the horizon phase structure. In contrast to the extremal boundary, which is obtained from the simultaneous conditions defining a degenerate horizon, the non-extremal boundary corresponds to a horizon reaching the spacetime origin. The latter can be obtained directly from the small-radius limit of the metric function $f(r=r_{\rm h})$. In the limit $r_{\rm h}\to 0$, Eq.~\eqref{eq:exact_fr} reduces to
	\begin{equation}\label{eq:non_extremal_condition}
		C(\epsilon,Q;M)=0.
	\end{equation}
	Using Eq.~\eqref{eq:norm_mass}, this condition yields the analytical non-extremal boundary
	\begin{align}\label{eq:non_extremal_curve}
		Q_{\rm nc}(\epsilon;M)
		=
		\gamma\,M^{2/3}\epsilon^{1/6},
	\end{align}
	where
	\begin{align}\label{eq:gamma}
		\gamma
		=
		\frac{3^{4/3}}
		{2^{11/6}\,
			\pi^{1/3}\,
			\Gamma\!\left(\tfrac14\right)^{4/3}}.
	\end{align}
	This curve can therefore be determined analytically without solving the horizon equation numerically.
	\begin{figure}[tb]
		\centering
		\begin{subfigure}{.5\textwidth}
			\centering
			\includegraphics[width=1\linewidth]{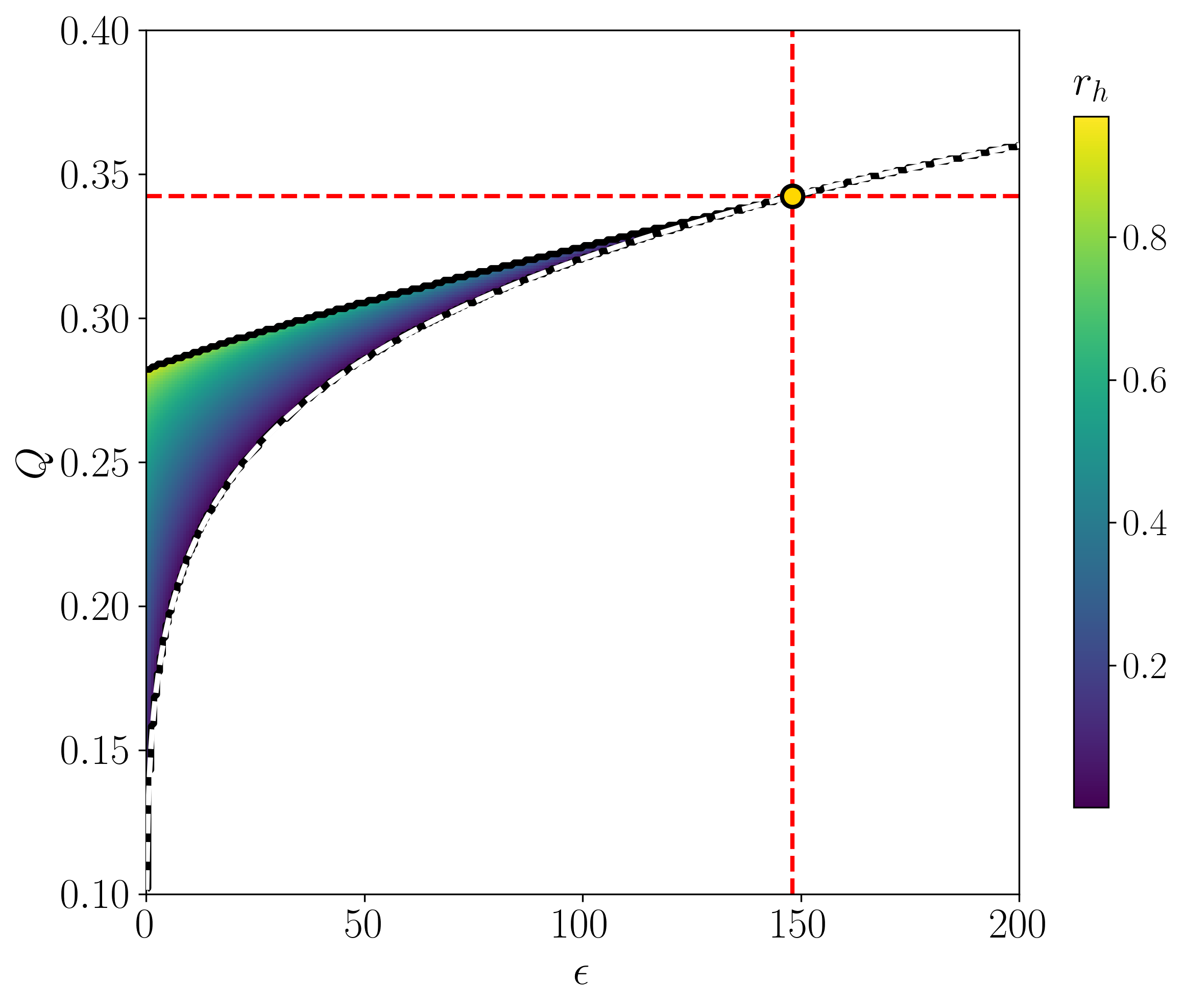}
		\end{subfigure}%
		\begin{subfigure}{.5\textwidth}
			\centering
			\includegraphics[width=1\linewidth]{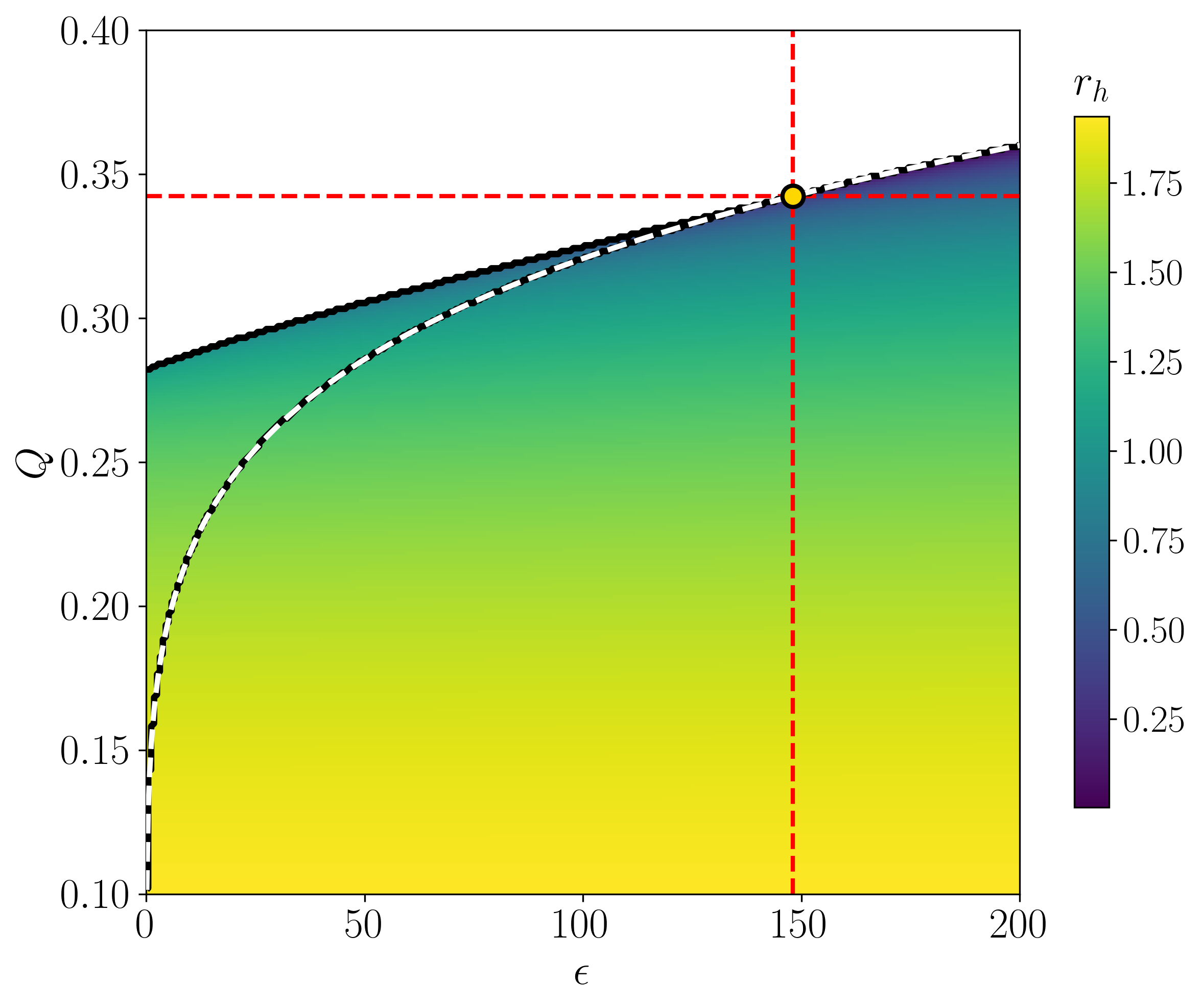}
		\end{subfigure}
		\caption{Numerical horizon radii mapped onto the $(Q,\epsilon)$ parameter space for $M=1.0$, $\kappa=8\pi$, and $\Lambda=0$. The left and right panels show the inner and outer horizon radii, respectively. The dashed white curve represents the analytical non-extremal boundary given by Eq.~\eqref{eq:non_extremal_curve}, while the black solid curves denote the numerically determined boundaries. The yellow circle marks the bifurcation point $(Q_{\rm b},\epsilon_{\rm b})$. For $\epsilon<\epsilon_{\rm b}$, the region between the non-extremal and extremal boundaries corresponds to black holes with two horizons. For $\epsilon>\epsilon_{\rm b}$, the two-horizon phase is absent, and the non-extremal boundary separates the single-horizon and horizonless phases.}
		\label{fig:rhoutinn_QepsM}
	\end{figure}

	The resulting horizon structure is shown in Fig.~\ref{fig:rhoutinn_QepsM}, where the numerical positive roots of $f(r_{\rm h})=0$ are mapped onto the $(Q,\epsilon)$ parameter space for $M=1.0$, $\kappa=8\pi$, and $\Lambda=0$. The left and right panels display the inner and outer horizon radii, respectively. The dashed white curve corresponds to the analytical non-extremal boundary given by Eq.~\eqref{eq:non_extremal_curve}, while the black solid curves denote the numerically determined boundaries. The agreement between the analytical curve and the corresponding numerical boundary provides a direct verification of the $r_{\rm h}\to0$ analysis. The upper black boundary corresponds to the finite-radius extremal branch $Q_{\rm ext}(\epsilon; M)$, while the yellow circle marks the bifurcation point ($Q_{\rm b} = 0.342334$, $\epsilon_{\rm b} = 148.050692$).

	The distribution of the horizon radii reveals the phase structure directly. For $\epsilon<\epsilon_{\rm b}$, the non-extremal and extremal boundaries enclose a region in which both inner and outer horizons are present. For $Q<Q_{\rm nc}$ (or $C > 0$), only a single outer horizon exists, while beyond the extremal curve no positive real root of $f(r)=0$ exists, and the configuration is horizonless. Thus, in this regime, the non-extremal boundary separates the single-horizon and two-horizon phases, while the extremal boundary separates the two-horizon and horizonless phases. The extremal boundary terminates at the bifurcation point where $r_{\rm h}^{\rm ext}\to0$, and at $r_{\rm h}^{\rm ext}=0$ both conditions $f(r_{\rm h}^{\rm ext}) = 0$ and $f^{\prime}(r_{\rm h}^{\rm ext}) = 0$ are satisfied, leaving a point-like configuration called an extremal black point with vanishing temperature and entropy ~\cite{Soleng1995}. At this point the non-extremal and extremal boundaries coincide. Indeed, using Eqs.~\eqref{eq:QM_cr} and ~\eqref{eq:eps_M_cr} in Eq.~\eqref{eq:non_extremal_curve} gives
	\begin{equation}\label{eq:Qnc_epsb}
		Q_{\rm nc}(\epsilon_{\rm b};M)=Q_{\rm b}(M),
	\end{equation}
	and hence
	\begin{equation}\label{eq:bifurcation_point}
		(r_{\rm h},Q,\epsilon)=(0,Q_{\rm b},\epsilon_{\rm b}).
	\end{equation}
	For $\epsilon > \epsilon_{\rm b}$, the finite-radius extremal branch no longer exists and the two-horizon phase disappears. The curve $C=0$ then separates 
	the single-horizon and horizonless configurations: the inner horizon is absent, while the outer horizon approaches $r_{\rm h} = 0$ on this boundary. Since 
	$\epsilon \neq \epsilon_{\rm b}$, we have $f(0) \neq 0$, so $r_{\rm h} = 0$ is not a horizon solution---the boundary itself is horizonless. Just before the 
	horizon vanishes, however, the configuration can be identified as an ordinary black point with zero entropy and diverging temperature~\cite{Sokolov2025}. Beyond this boundary ($Q > Q_{\rm nc}$), the outer horizon disappears entirely and the configuration becomes horizonless.

	\subsection{Thermodynamic Analysis of Black Points} \label{subsec:black_points}
	
	To understand why the Soleng-type extremal black point has a vanishing temperature while ordinary black points have a diverging one, it is convenient to assume that a horizon solution of $f(r)=0$ exists and then expand $f(r)$ in Eq.~\eqref{eq:exact_fr} for small $r$:
	\begin{align}\label{eq:smallr_fr}
		f(r) \sim
		-
		\frac{2C(\epsilon, Q; M)}{r}
		+
		f_{0}
		+ \cdots
		,
	\end{align}
	with radial derivative
	\begin{align}\label{eq:smallr_frprime}
		f^{\prime}(r)
		\sim
		\frac{2C(\epsilon, Q; M)}{r^{2}}
		+ \cdots
		,
	\end{align}
	where
	\begin{align}
		f_{0} 
		=
		1 - \frac{\sqrt{2}\kappa |Q|}{\sqrt{\epsilon}}
		.
	\end{align}
	Imposing the horizon condition $f(r=r_{\rm h})=0$ in Eq.~\eqref{eq:smallr_fr} then fixes the scaling of $C$ in terms of $r_{\rm h}$,
	\begin{align}
		C
		\sim
		\frac{f_{0}}{2}
		r_{\rm h},
	\end{align}
	and substituting this back into Eq.~\eqref{eq:smallr_frprime} gives
	\begin{align}
		f^{\prime}(r_{\rm h}) \sim
		\frac{f_{0}}{r_{\rm h}}.
	\end{align}
	The corresponding Hawking temperature, entropy, and heat capacity are therefore
	\begin{align}\label{eq:temperature}
		T_{\rm 0}
		=
		\frac{f'(r_{\rm h})}{4\pi}
		\sim
		\frac{f_{0}}{4 \pi r_{\rm h}},
	\end{align}
	\begin{align}\label{eq:entropy}
		S_{\rm 0}  = \pi r_{\rm h}^{2},
	\end{align}
	and
	\begin{align}\label{eq:heat_capacity}
		C_{\rm 0}
		&=
		T_{\rm 0}\,
		\frac{\partial S_{\rm 0}}
		{\partial T_{\rm 0}}
		=
		T_{\rm 0}\,
		\frac{\partial S_{\rm 0}/\partial r_{\rm h}}
		{\partial T_{\rm 0}/\partial r_{\rm h}}
		\nonumber \\
		&=
		2\pi r_{\rm h} \frac{f^{\prime}(r_{\rm h})}{f^{\prime\prime}(r_{\rm h})}
		\sim
		- 2 \pi r_{\rm h}^{2}.
	\end{align}
	With these scalings in hand, the two types of black points are easy to distinguish. At $\epsilon = \epsilon_{\rm b}$, as we increase $Q$ we hit the bifurcation point $Q = Q_{\rm nc} = Q_{\rm b}$, which we identify as the Soleng-type black point. At this point $f_{0}$ vanishes, and hence $f^{\prime} \to 0$ as $r_{\rm h} \to 0$, so all thermodynamic quantities vanish at $r_{\rm h}=0$. For ordinary black points, however, $\epsilon > \epsilon_{\rm b}$, so $f_{0}$ remains finite and positive as $Q \to Q_{\rm nc}$. Therefore $f^{\prime} \to \infty$ as $r_{\rm h} \to 0$, and in turn $T_{\rm 0} \to \infty$, while both $S_{\rm 0}$ and $C_{\rm 0}$ vanish. This confirms that, in contrast to the Soleng-type black points with vanishing temperature, the ordinary black points have a diverging temperature.
	
	Furthermore, we note that for ordinary black points, as $r_{\rm h}\to0$, the entropy and heat capacity take the Schwarzschild forms $S_{0}=\pi r_{\rm h}^{2}$ and $C_{0}=-2\pi r_{\rm h}^{2}$, while the temperature retains the prefactor $f_{0}=1-\sqrt{2}\kappa|Q|/\sqrt{\epsilon}$, so that $T_{0}=f_{0}/(4\pi r_{\rm h})$ differs from the Schwarzschild temperature by exactly this factor. The black-point boundary ties the charge to the mass through $Q=Q_{\rm nc}=\gamma M^{2/3}\epsilon^{1/6}$, so $f_{0}$ cannot be varied independently: it approaches unity only in the small-mass limit, where the charge contribution to $f_{0}$ becomes negligible. In that limit the temperature also reduces to the Schwarzschild form, and the ordinary LNED black point tends smoothly to the Schwarzschild black point as $M\to0$. For Soleng-type black points, by contrast, $f_{0}=0$ identically, so the temperature vanishes at $r_{\rm h}\to0$ and no Schwarzschild correspondence arises.
	
	These two types of black points fit within the broader classification established for gravitating nonlinear electrodynamics. In general NED models, the thermodynamic behaviour of black points at vanishing horizon radius depends on the structure of the electrostatic field near the origin: for a wide class of Lagrangians, the temperature diverges for non-extremal black points and either vanishes, diverges, or takes a finite universal value for extremal black points depending on the model parameters~\cite{DiazAlonso2013}. The ordinary black point found here, with $T \to \infty$ and $S \to 0$, falls into the non-extremal category, while the Soleng-type configuration corresponds to the extremal case with $T \to 0$ and $S \to 0$~\cite{Soleng1995}. More recent work has extended this classification to vacuum NED, where the third law is shown to be fulfilled for black points~\cite{Sokolov2025}.
	
	Within logarithmic NED specifically, the combined effects of nonlinear electrodynamics and quantum entropy corrections remain largely unexplored. A recent study incorporating exponential entropy corrections~\cite{Brzo2026} considered the modified entropy
	\begin{equation}
		S_{\eta}(r_{\rm h}) = S_{0}(r_{\rm h}) + \eta\,e^{-\delta S_{0}(r_{\rm h})},
		\qquad
		S_{0}(r_{\rm h}) = \pi r_{\rm h}^{2},
	\end{equation}
	where $S_{0}$ is the classical Bekenstein--Hawking entropy and $\eta$, $\delta$ are the parameters controlling the non-perturbative correction. This correction yields a well-defined, non-zero entropy at the origin,
	\begin{equation}
		\lim_{r_{\rm h} \to 0} S_{\eta}(r_{\rm h}) = \eta.
	\end{equation}
	When the temperature is derived from the thermodynamic first law $\mathrm{d}M = T\,\mathrm{d}S$ with the corrected entropy, the correction propagates into the temperature through the factor
	\begin{equation}
		H_{\rm \eta}(r_{\rm h}) = \frac{1}{2\pi r_{\rm h}}\frac{\mathrm{d}S_{\eta}}{\mathrm{d}r_{\rm h}}
		= 1 - \eta\delta\,e^{-\delta\pi r_{\rm h}^{2}},
	\end{equation}
	yielding
	\begin{equation}
		T_{\rm \eta} = \frac{T_{0}}{H_{\rm \eta}},
		\qquad
		T_{0} = \frac{f'(r_{\rm h})}{4\pi}.
	\end{equation}
	The corresponding heat capacity is
	\begin{align}
		C_{\rm \eta} =
		\frac{
			H_{\rm\eta} \; C_{\rm 0} \; \partial M / \partial r_{\rm h}
		}
		{\partial M / \partial r_{\rm h}
			- C_{\rm 0} \; T_{\rm \eta} \; \partial H_{\rm \eta} / \partial r_{\rm h}
		},
	\qquad
	C_{0} = 2\pi r_{\rm h} \frac{f^{\prime}(r_{\rm h})}{f^{\prime\prime}(r_{\rm h})}.
	\end{align}
	Where, from Eq.~\eqref{eq:f_r} and assuming $f(r=r_{\rm h})=0$,
	\begin{align}
		\frac{\partial M}{\partial r_{\rm h}} = \frac{1}{2} - \frac{\Lambda  r_{\rm h}^{2}}{2}  -\frac{\kappa}{2} r_{\rm h}^{2}\rho(r_{\rm h}).
	\end{align}
	For $\eta=0$, we have $H_{\rm \eta}=1$ and $\partial H_{\rm \eta}/\partial r_{\rm h} = 0$, and the above thermodynamic quantities reduce to their uncorrected counterparts $S_{\rm 0}$, $T_{\rm 0}$, and $C_{\rm 0}$.
	
	In the limit $r_{\rm h} \to 0$, the corrected entropy behaves as
	\begin{align}
		S_{\rm \eta} \sim \eta +
		(1-\eta\delta) \pi r_{\rm h}^{2} +
		\frac{1}{2} \eta\delta^{2} \pi^{2} r_{\rm h}^{4},
	\end{align}
	so that $S_{\eta} \to \eta$ as $r_{\rm h} \to 0$. The temperature and heat capacity, however, exhibit two distinct asymptotic regimes depending on whether $\eta\delta = 1$ or not. For $\eta\delta \neq 1$,
	\begin{align}
		T_{\rm\eta} \sim
		\frac{1}{1-\eta\delta} \;
		\frac{f_{0}}{4\pi r_{\rm h}}
		+ \mathcal{O}(r_{\rm h} \, \ln r_{\rm h})
		,
		\\
		C_{\rm\eta} \sim -2\big(1-\eta\delta\big)^{2}\pi r_{\rm h}^{2}
		+ \mathcal{O}(r_{\rm h}^{4} \, \ln r_{\rm h})
		.
	\end{align}
	For the extremal Soleng-type black points, where $f_{0}=0$, both $T_{\eta} \to 0$ and $C_{\eta} \to 0$ as $r_{\rm h} \to 0$. For ordinary black points, where $f_{0}\neq 0$, we have $C_{\eta} \to 0$ but $T_{\eta} \to \pm\infty$, the sign being determined by $1-\eta\delta$; we restrict to $\eta\delta\le 1$, for which $T_{\eta}\to+\infty$ and the temperature remains non-negative. For the special case $\eta\delta=1$,
	\begin{align}
		T_{\rm\eta} \sim
		\frac{f_{0}}{4\delta\pi^{2} r_{\rm h}^{3}}
		- \frac{\kappa}{2\delta\pi^{2}\epsilon}
		\frac{\ln r_{\rm h}}{r_{\rm h}}
		,
		\\
		C_{\rm\eta} \sim
		-\frac{2\delta\pi^{2}}{3} r_{\rm h}^{4}
		+ \mathcal{O}(r_{\rm h}^{6} \, \ln r_{\rm h})
		.
	\end{align}
	Here the extremal case behaves differently: even for $f_{0}=0$, $T_{\rm \eta}$ no longer vanishes but diverges to $+\infty$ as $\ln r_{\rm h}/r_{\rm h}$, qualitatively resembling an ordinary black point, though with a logarithmic enhancement rather than the power-law divergence $\sim 1/r_{\rm h}$ found for $f_{0}\neq 0$, while $C_{\rm \eta}$ still vanishes.

	In summary, the quantum correction does yield black points with non-zero entropy, but it does not alter the qualitative behavior of $T_{\rm \eta}$ and $C_{\rm \eta}$ at $r_{\rm h} \to 0$ for $\eta\delta \neq 1$: ordinary black points still exhibit a diverging temperature and vanishing heat capacity, while extremal Soleng-type black points still exhibit vanishing temperature and heat capacity. The functional dependence on $r_{\rm h}$, however, is modified by the correction. For $\eta\delta \neq 1$, the powers of $r_{\rm h}$ are unchanged, but the coefficients are rescaled by factors of $(1-\eta\delta)$. The special case $\eta\delta = 1$ is different: the powers themselves change, and even the extremal black point no longer has a vanishing temperature. Instead, $T_{\rm \eta} \sim - \ln r_{\rm h}/r_{\rm h}$ diverges, while $C_{\rm \eta} \sim r_{\rm h}^{4}$, in contrast to the uncorrected $1/r_{\rm h}$ and $r_{\rm h}^{2}$. Moreover, in the absence of backreaction, the geometric Hawking temperature $T_{\rm 0} = f'(r_{\rm h})/(4\pi)$ remains unchanged, and the thermodynamic temperature $T_{\rm \eta}$ does not coincide with it. Whether the inclusion of backreaction would reconcile the two temperatures, or whether the phenomenological $T_{\rm \eta}$ captures the leading correction, thus remains an open problem.	
		
	Finally, having established the horizon phase structure, we now turn to the null-geodesic structure of the background spacetime, focusing on the circular null orbits, their merger structure, and the associated effective potential.

\section{Background Null-Geodesic Structure}\label{sec:PMBG}

As discussed in the previous sections, the logarithmic nonlinearity modifies the spacetime geometry through the metric function $f(r)$. We now investigate the motion of massless test particles following null geodesics of the background spacetime. The Lagrangian for geodesic motion is
\begin{align}\label{eq:Lag_density0}
	\mathcal{L} = \frac{1}{2} g_{\mu\nu}\dot{x}^{\mu}\dot{x}^{\nu},
\end{align}
where the dot denotes differentiation with respect to an affine parameter $\tau$. Substituting the metric coefficients from~\eqref{eq:metric}, we obtain
\begin{align}\label{eq:Lag_density1}
	\mathcal{L} = \frac{1}{2} \Bigg[-f(r)\dot{t}^{2} + \frac{\dot{r}^{2}}{f(r)} + r^{2}\dot{\theta}^{2} + r^{2} \sin^{2}(\theta)\dot{\phi}^{2}\Bigg].
\end{align}
By spherical symmetry, the motion can be restricted to the equatorial plane, $\theta=\pi/2$, with $\dot{\theta}=0$. The Lagrangian then reduces to
\begin{align}\label{eq:Lag_density2}
	\mathcal{L} = \frac{1}{2} \Bigg[-f(r)\dot{t}^{2} + \frac{\dot{r}^{2}}{f(r)} + r^{2} \dot{\phi}^{2}\Bigg].
\end{align}

Since $t$ and $\phi$ are cyclic coordinates, the corresponding conserved quantities are the energy $E$ and angular momentum $L$,
\begin{align}\label{eq:conserved_quantities}
	E=-\frac{\partial\mathcal{L}}{\partial\dot{t}}=f(r)\dot{t},
	\qquad
	L=\frac{\partial\mathcal{L}}{\partial\dot{\phi}}=r^{2}\dot{\phi},
\end{align}
which give
\begin{align}\label{eq:cons_E}
	\dot{t} = \frac{E}{f(r)}
\end{align}
and
\begin{align}\label{eq:cons_L}
	\dot{\phi} = \frac{L}{r^{2}}.
\end{align}
For massless test particles, the geodesics satisfy the null condition $ds^{2}=0$, or equivalently $2\mathcal{L}=0$. Substituting Eqs.~\eqref{eq:cons_E} and~\eqref{eq:cons_L} into~\eqref{eq:Lag_density2} yields
\begin{align}\label{eq:total_E}
	\dot{r}^{2} + V_{\rm eff}^{\rm bg}(r) = E^{2},
\end{align}
where the background effective potential is
\begin{align}\label{eq:Veff_bg_r}
	V_{\rm eff}^{\rm bg}(r) = L^{2}\frac{f(r)}{r^{2}}.
\end{align}
Circular null orbits are characterized by constant radius $r=r_{c}$ and therefore satisfy the stationary condition $\frac{dV_{\rm eff}^{\rm bg}(r)}{dr}\Big. \Big|_{r=r_{c}} = 0$, leading to
\begin{align}\label{eq:bg_circular_condition}
	r_{\rm c}\frac{df}{dr}\bigg|_{r=r_{\rm c}}-2f(r_{\rm c})=0.
\end{align}
The same circular-orbit condition applies to null geodesics in non-rotating charged Reissner-Nordstr\"om black-hole spacetimes~\cite{Wei2020}. However, in the rotating charged case, the interplay between charge and rotation produces a richer and qualitatively different circular-orbit structure, as illustrated for extremal Kerr--Newman black holes~\cite{Chen2025}.

Combining Eq.~\eqref{eq:bg_circular_condition} with Eq.~\eqref{eq:diff_f} eliminates the derivative of the metric function and gives
\begin{align}\label{eq:bg_circular_condition_2}
	3f(r_{\rm c})=1-\Lambda r_{\rm c}^2-\kappa r_{\rm c}^2\rho(r_{\rm c}).
\end{align}
Substituting $E(r_{\rm c})$, $\rho(r_{\rm c})$, and $f(r_{\rm c})$ from Eqs.~\eqref{eq:electric_field}, \eqref{eq:rho}, and \eqref{eq:exact_fr}, respectively, we obtain
\begin{align}\label{eq:radius_circle}
	F(r_{\rm c};\epsilon, Q, M) &= r_{\rm c} - 3 C(\epsilon, Q; M) 
	- 
	\frac{\kappa}{3\epsilon} r_{\rm c} \Big(\sqrt{r_{\rm c}^{4} + 2\epsilon Q^{2}} - r_{\rm c}^{2}\Big) 
	\\ \nonumber
	&- \sqrt{\frac{8}{\epsilon}}\frac{|Q|\kappa}{3} r_{\rm c}
	\left[ {}_2F_1\left(\tfrac{1}{4},\tfrac{1}{2};\tfrac{5}{4};
	-\frac{r_{\rm c}^{4}}{2\epsilon Q^2}\right)\right]=0.
\end{align}
Here, $C(\epsilon,Q;M)$ is related to the mass $M$ through Eq.~\eqref{eq:norm_mass}. Because Eq.~\eqref{eq:radius_circle} contains a hypergeometric function, its general solution is not available in closed form, and the circular-orbit radii must in general be determined numerically.
\begin{figure}[t!]
	\centering
	\begin{subfigure}{0.49\textwidth}
		\centering
		\includegraphics[width=\linewidth]{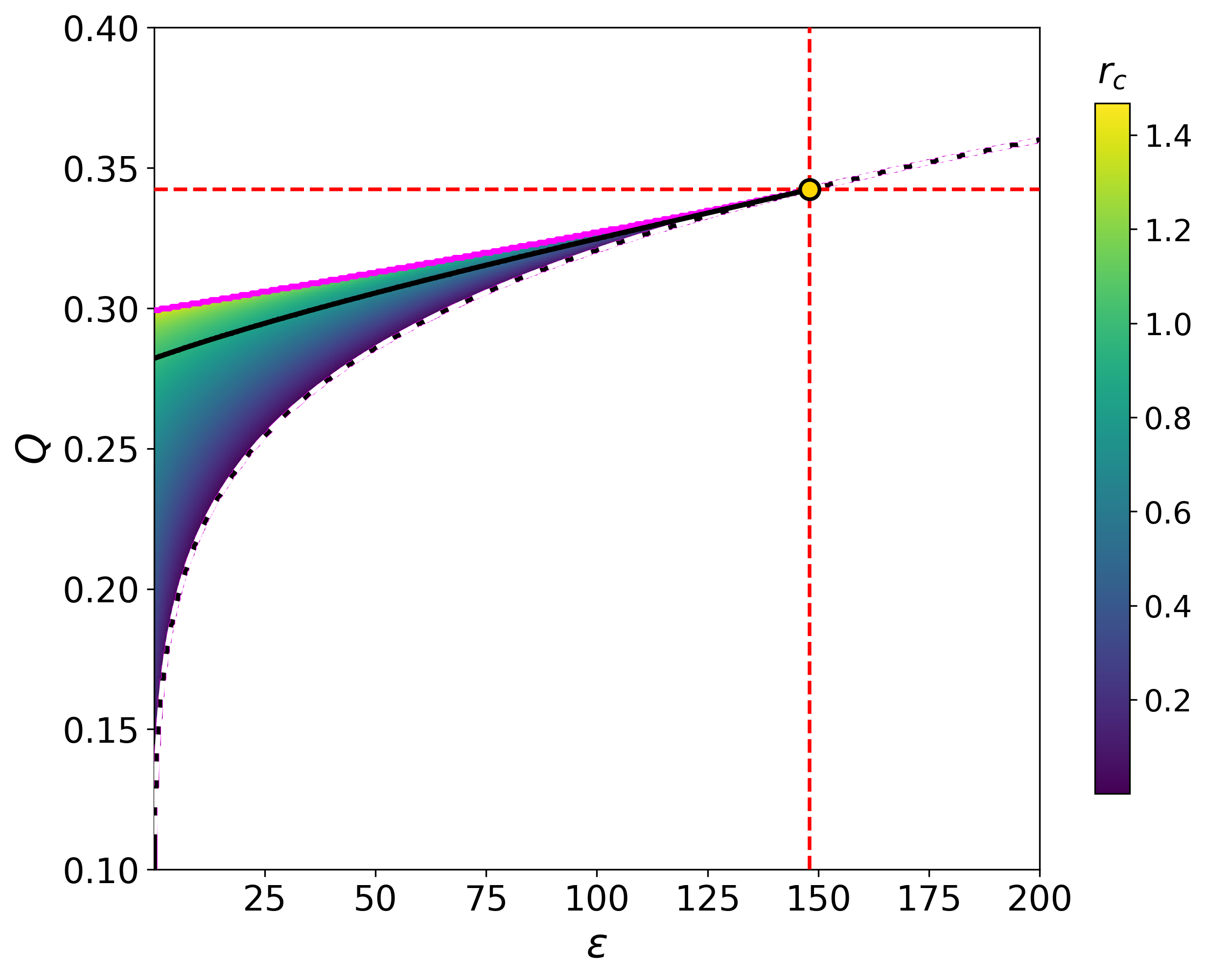}
		\caption{\footnotesize{Inner stable circular orbits}} 	
	\end{subfigure}
	\hfill
	\begin{subfigure}{0.49\textwidth}
		\centering
		\includegraphics[width=\linewidth]{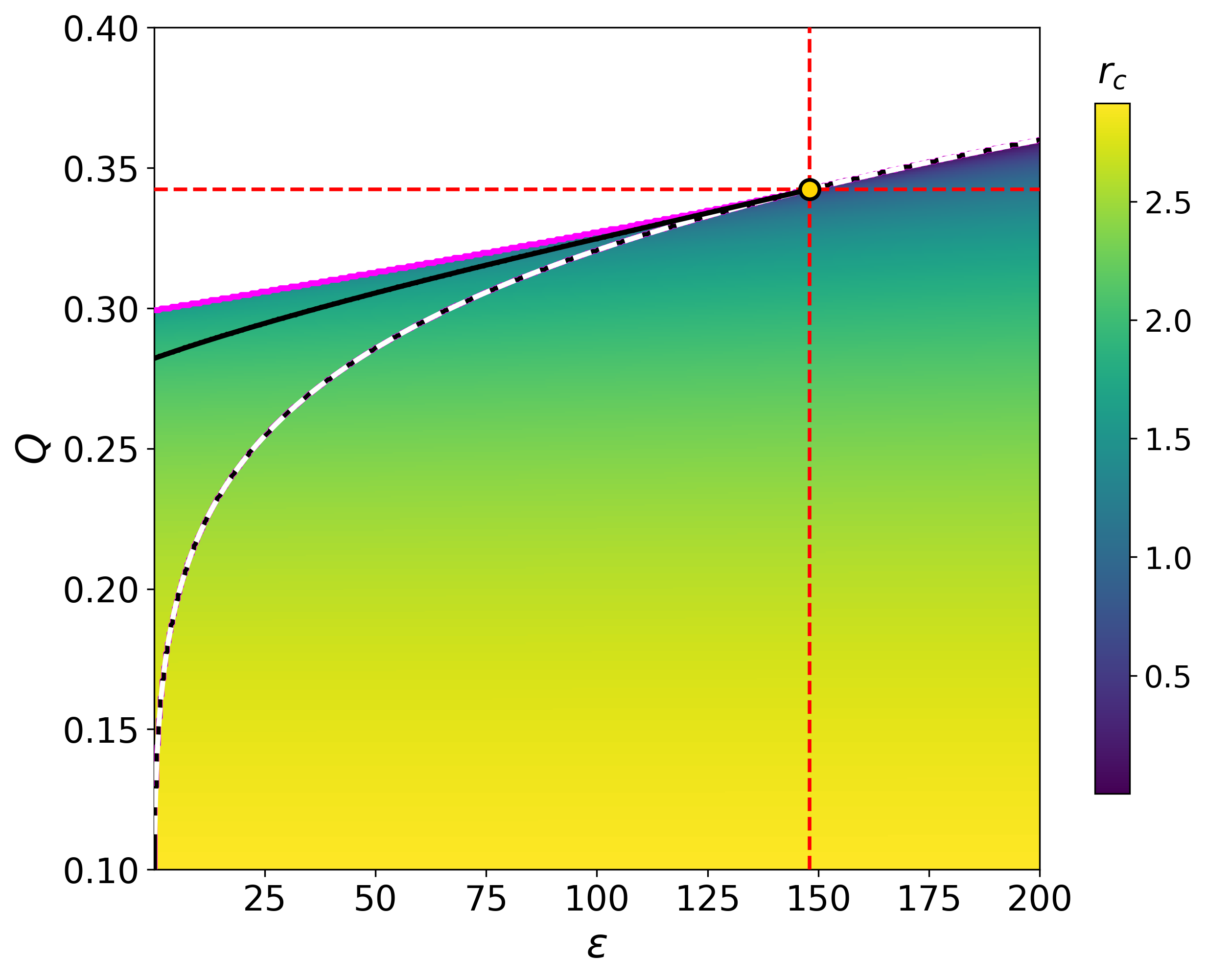}
		\caption{\footnotesize{Outer unstable circular orbits}}   	
	\end{subfigure}
	
	\vspace{0.5cm}
	
	\begin{subfigure}{0.49\textwidth}
		\centering
		\includegraphics[width=\linewidth]{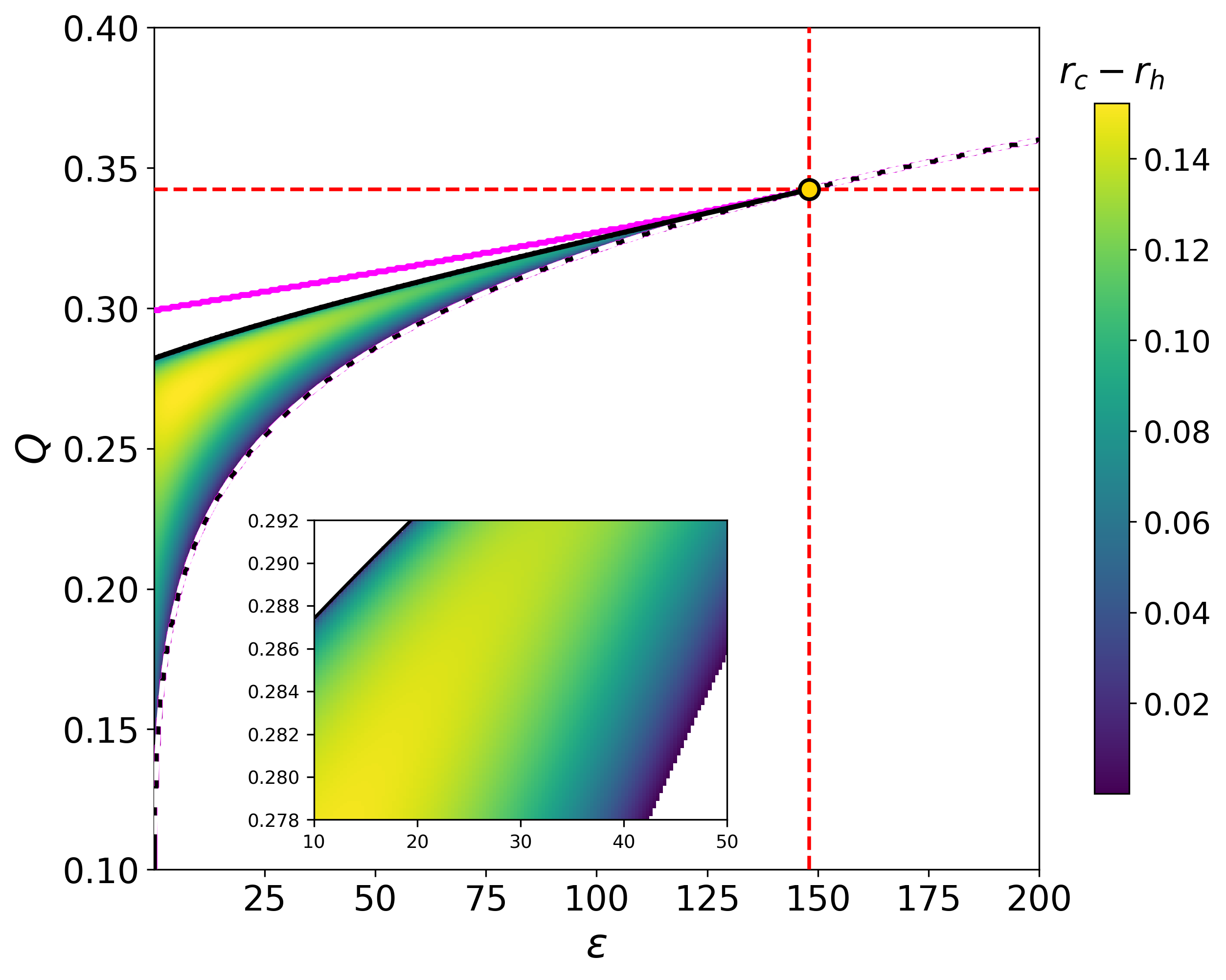}
		\caption{\footnotesize{Inner orbit-horizon separation}}  	
	\end{subfigure}
	\hfill
	\begin{subfigure}{0.49\textwidth}
		\centering
		\includegraphics[width=\linewidth]{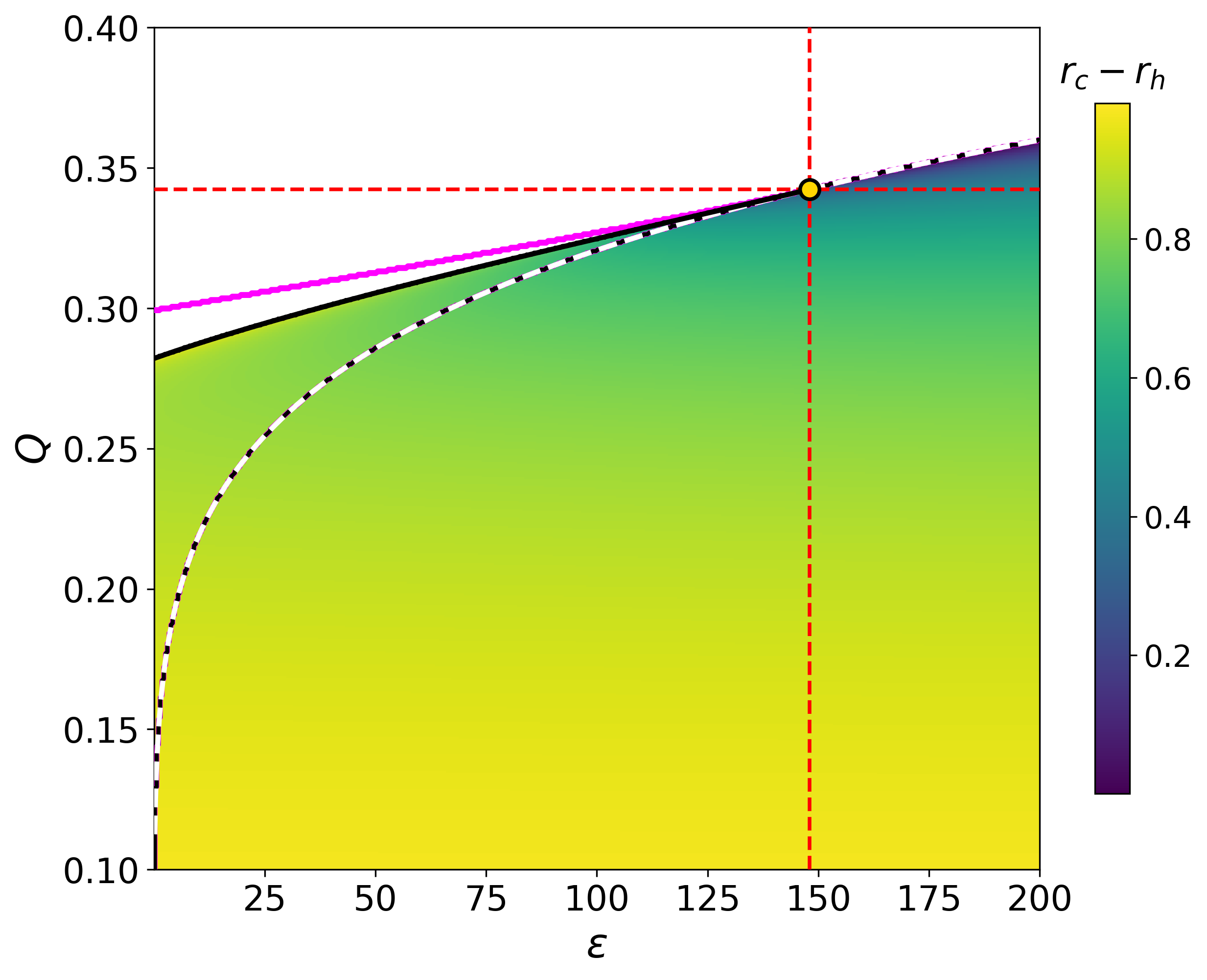}
		\caption{\footnotesize{Outer orbit-horizon separation}}
	\end{subfigure}
	\caption{Circular-orbit radii and orbit--horizon separations mapped onto the $(Q,\epsilon)$ parameter space for $M=1.0$, $\kappa=8\pi$, and $\Lambda=0$. The upper panels show the radii of the inner stable and outer unstable circular orbits, respectively, while the lower panels display the corresponding orbit--horizon separations, $r_{\rm c}-r_{\rm h}$. The critical circular-orbit boundary is represented by the blue curve, while the numerical and analytical horizon boundaries are represented by the solid black and dashed white curves, respectively. The yellow circle marks the common bifurcation point $(Q_{\rm b},\epsilon_{\rm b})$. The critical circular-orbit boundary extends into the horizonless side of the extremal horizon boundary, indicating a finite region in which both stable and unstable circular null orbits coexist without an event horizon.}
	\label{fig:rcrh_diff}
\end{figure}

The resulting circular-orbit structure is shown in Fig.~\ref{fig:rcrh_diff}, where the positive roots of Eq.~\eqref{eq:radius_circle} are mapped onto the $(Q,\epsilon)$ parameter space for $M=1.0$, $\kappa=8\pi$, and $\Lambda=0$. The upper panels display the radii of the inner and outer circular orbits. The two circular-orbit branches have different stability properties: the inner branch corresponds to a local minimum of $V_{\rm eff}^{\rm bg}(r)$ and is stable against radial perturbations, whereas the outer branch corresponds to a local maximum and is unstable. These features are illustrated by the effective-potential plots in Subsec.~\ref{subsec:Veff_bg}. They merge at a critical boundary, beyond which no finite positive circular null orbit of the background metric exists. However, the lower panels of Fig.~\ref{fig:rcrh_diff} show the corresponding orbit--horizon separations, $r_{\rm c}-r_{\rm h}$, providing a direct comparison between the circular-orbit and horizon locations. Wherever horizons are present, the plotted separations remain non-negative, indicating that the circular-orbit branches shown in the figure are not hidden behind the corresponding horizon branches. Together with the upper panels, these results show that the circular-orbit structure can persist even after the horizon structure has entered the horizonless regime.

Furthermore, a non-critical boundary of the circular-orbit structure follows directly from the limit $r_{\rm c}\rightarrow0$ of Eq.~\eqref{eq:radius_circle}. In this limit, the hypergeometric and power-law contributions vanish, leaving
\begin{equation}\label{eq:non_critical_condition}
	C(\epsilon,Q;M)=0,
\end{equation}
which is precisely the condition defining the non-extremal horizon boundary obtained in the previous section. Hence, the circular-orbit boundary associated with $r_{\rm c}\rightarrow0$ coincides analytically with the non-extremal horizon boundary. This correspondence follows from the common condition $C(\epsilon,Q;M)=0$ and is therefore not merely a numerical coincidence. As shown in Fig.~\ref{fig:rcrh_diff}, the circular-orbit structure parallels that of the horizon. For $\epsilon \ge \epsilon_{b}$, where the black points reside, both retain only their outer branch, the inner branch being absent in each case. On this segment of the boundary, both the outer horizon and the outer circular orbit shrink to $r=0$, so the black points admit no circular orbits of finite radius.

The second boundary is determined by the coalescence of the two circular-orbit branches. It is obtained from the simultaneous conditions
\begin{align}\label{eq:merger_conditions}
	F(r_{\rm c};\epsilon,Q,M)=0,
	\qquad
	\left.\frac{\partial F(r;\epsilon,Q,M)}{\partial r}\right|_{r=r_{\rm c}}=0.
\end{align}
with $r_{\rm c}$ finite in general. At this locus, the stable and unstable circular orbits merge into a single marginally stable orbit, corresponding to a double root of the circular-orbit equation and a degenerate stationary point of the effective potential. In the $(Q,\epsilon)$ parameter space, this critical circular-orbit boundary lies on the horizonless side of the extremal horizon boundary. Consequently, a finite wedge of parameter space supports both stable and unstable circular null orbits without an event horizon. This boundary is shown as the blue curve in Fig.~\ref{fig:rcrh_diff}. Such horizonless circular-orbit configurations have also been found in other regular black-hole and nonlinear-electrodynamics models~\cite{Stuchlik2015,Chiba2017,Toshmatov2026}. The common bifurcation point of the horizon and circular-orbit structures can be obtained analytically by differentiating Eq.~\eqref{eq:radius_circle} with respect to $r_{\rm c}$, which gives
\begin{align}\label{eq:radius_circle_1drv}
	1
	-
	\frac{\kappa}{\epsilon}
	\left(
	\sqrt{r_{\rm c}^{4}+2\epsilon Q^{2}}
	-
	r_{\rm c}^{2}
	\right)=0,
\end{align}
where Eq.~\eqref{eq:hypergeometric_integral} has been used to simplify the derivative of the hypergeometric contribution. At the bifurcation point, the marginal circular orbit reaches the spacetime origin, $r_{\rm c}\rightarrow0$. Evaluating Eqs.~\eqref{eq:radius_circle} and~\eqref{eq:radius_circle_1drv} in this limit yields the same two conditions that lead to Eqs.~\eqref{eq:QM_cr} and ~\eqref{eq:eps_M_cr}. Thus, the circular-orbit and horizon phase structures share the same bifurcation point $(Q_{\rm b},\epsilon_{\rm b})$.

Equation~\eqref{eq:radius_circle_1drv} can be solved exactly for the radius of the marginally stable circular orbit to obtain:
\begin{align}\label{eq:rc_crit}
	r_{\rm m} =
	\sqrt{\kappa} \big|Q_{\rm m}(\epsilon; M)\big|
	\left(1-\frac{\epsilon}{\epsilon_{\rm m}}\right)^{1/2},
\end{align}
provided that $\epsilon\leq\epsilon_{\rm m}$, where $\kappa=8\pi$ and
\begin{align}\label{eq:epsilon_m}
	\epsilon_{\rm m}
	&=2\kappa^{2}\big|Q_{\rm m}(\epsilon;M)\big|^{2} \nonumber\\
	&=128\pi^{2}\big|Q_{\rm m}(\epsilon;M)\big|^{2}.
\end{align}
Here, $r_{\rm m}$ denotes the radius of the marginally stable circular orbit and $Q_{\rm m}(\epsilon;M)$ is the corresponding critical charge along the critical circular-orbit boundary. Substituting Eq.~\eqref{eq:rc_crit} into Eq.~\eqref{eq:radius_circle} gives the implicit master equation
\begin{align}\label{eq:Fc_crit}
	F_{\rm m}(\epsilon, Q_{\rm m}; M)
	={}&
	1 - \frac{3C(\epsilon, Q_{\rm m}; M)}{\sqrt{\kappa}\,|Q_{\rm m}|\sqrt{\delta}}
	-\frac{\kappa}{3\epsilon}
	\left[
	|Q_{\rm m}|
	\sqrt{\kappa^2Q_{\rm m}^2\delta^2+2\epsilon}
	-\kappa Q_{\rm m}^2\delta
	\right]
	\nonumber\\
	&-
	\frac{\kappa|Q_{\rm m}|}{3}
	\sqrt{\frac{8}{\epsilon}}\,
	{}_2F_1\!\left(
	\frac14,\frac12;
	\frac54;
	-\frac{\kappa^2Q_{\rm m}^2\delta^2}{2\epsilon}
	\right)=0,
\end{align}
where
\begin{align}\label{eq:delta}
	\delta = 1-\frac{\epsilon}{\epsilon_{\rm m}},
\end{align}
and $C(\epsilon,Q_{\rm m};M)$ is defined through Eq.~\eqref{eq:norm_mass}. Solving Eq.~\eqref{eq:Fc_crit} numerically determines the critical curve $Q_{\rm m}(\epsilon;M)$, depicted as the blue curve in the parameter map of Fig.~\ref{fig:rcrh_diff}. Substituting these numerical roots into Eq.~\eqref{eq:rc_crit} then determines the corresponding marginal radius $r_{\rm m}$. This hybrid analytical--numerical procedure avoids solving Eq.~\eqref{eq:radius_circle} and its radial derivative simultaneously, while retaining the exact analytic relation between $r_{\rm m}$, $Q_{\rm m}$, and $\epsilon$.

The resulting picture is therefore governed by two distinct boundary mechanisms: the non-critical boundary coincides with the non-extremal horizon boundary, whereas the critical merger boundary is displaced into the horizonless region and remains distinct from the extremal horizon boundary, the two meeting only at the common bifurcation point $(Q_{\rm b},\epsilon_{\rm b})$. This distinction highlights the nontrivial interplay between the LNED-modified geometry, its horizon structure, and the dynamics of background null geodesics.

	\subsection{Local Behavior of the Critical Merger Curve}\label{subsec:local_merger}
	
	To obtain analytical insight into the critical merger branch, we expand the implicit master equation~\eqref{eq:Fc_crit} in powers of $\epsilon$ about the Maxwell limit. Collecting terms order by order gives the local expression for the merger charge,
	\begin{align}\label{eq:Qm_extremal}
		Q_{\rm m}^2(\epsilon; M)
		=
		\frac{9M^2}{4\kappa}
		+\frac{\epsilon}{10\kappa^2}
		+\frac{4\epsilon^2}{2025\kappa^3M^2}
		+\mathcal O(\epsilon^3).
	\end{align}
	This expansion provides a direct analytical description of the critical charge near the RN limit and, as shown below, remains accurate well beyond the strict small-$\epsilon$ regime. Rather than expanding the merger radius independently, it is more useful to retain the exact algebraic relation obtained from the marginality condition~\eqref{eq:rc_crit},
	\begin{align}\label{eq:exact_numeric_rm}
		r_{\rm m} (\epsilon; M) = \frac{3M}{2} \sqrt{1 - \frac{8\epsilon}{45\kappa M^2} + \frac{16\epsilon^2}{18225\kappa^2 M^4} + \mathcal O(\epsilon^3)}.
	\end{align}
	Equation~\eqref{eq:exact_numeric_rm} is a hybrid approximation: the critical charge entering the exact relation for $r_{\rm m}$ is replaced by its perturbative expansion~\eqref{eq:Qm_extremal}, while the nonlinear dependence of the merger radius on this charge is retained in unexpanded form. This is advantageous because the merger radius is not uniformly well described by a straightforward Taylor expansion over the entire physical branch. As the bifurcation point is approached, $r_{\rm m}$ decreases rapidly and eventually vanishes, making a low-order polynomial expansion in $\epsilon$ increasingly sensitive to higher-order contributions. The hybrid expression instead preserves the square-root structure responsible for the termination of the physical branch.
	
	The RN limit provides a consistency check. At $\epsilon=0$, the marginally stable circular orbit reduces to the Reissner--Nordstr\"om result,
	\begin{align}\label{eq:RN_Q0}
		Q_{0} = 
		Q_{\rm m}(0,M) = \frac{3M}{2\sqrt{\kappa}},
	\end{align}
	and
	\begin{align}\label{eq:RN_R0}
		R_{0} = 
		r_{\rm m}(0,M)
		&= \frac{3M}{2}
		=\sqrt{\kappa}\,|Q_{\rm m}(0;M)|
		\nonumber \\
		&=\sqrt{2}\,|q_{\rm m}(0;M)|,
	\end{align}
	where $q=\sqrt{4\pi}\,Q$ denotes the conventional RN charge parameter. Thus, the merger occurs at $r_{\rm m}=3M/2$ and $Q_{\rm m}=3M/(2\sqrt{\kappa})$, corresponding to the circular-null-orbit merger in the overcharged RN regime, where no event horizon is present. The nonlinear theory therefore provides a continuous deformation of this RN critical configuration.
\begin{figure}[tb]
	\centering
	\begin{subfigure}{.49\textwidth}
		\centering
		\includegraphics[width=1\linewidth]{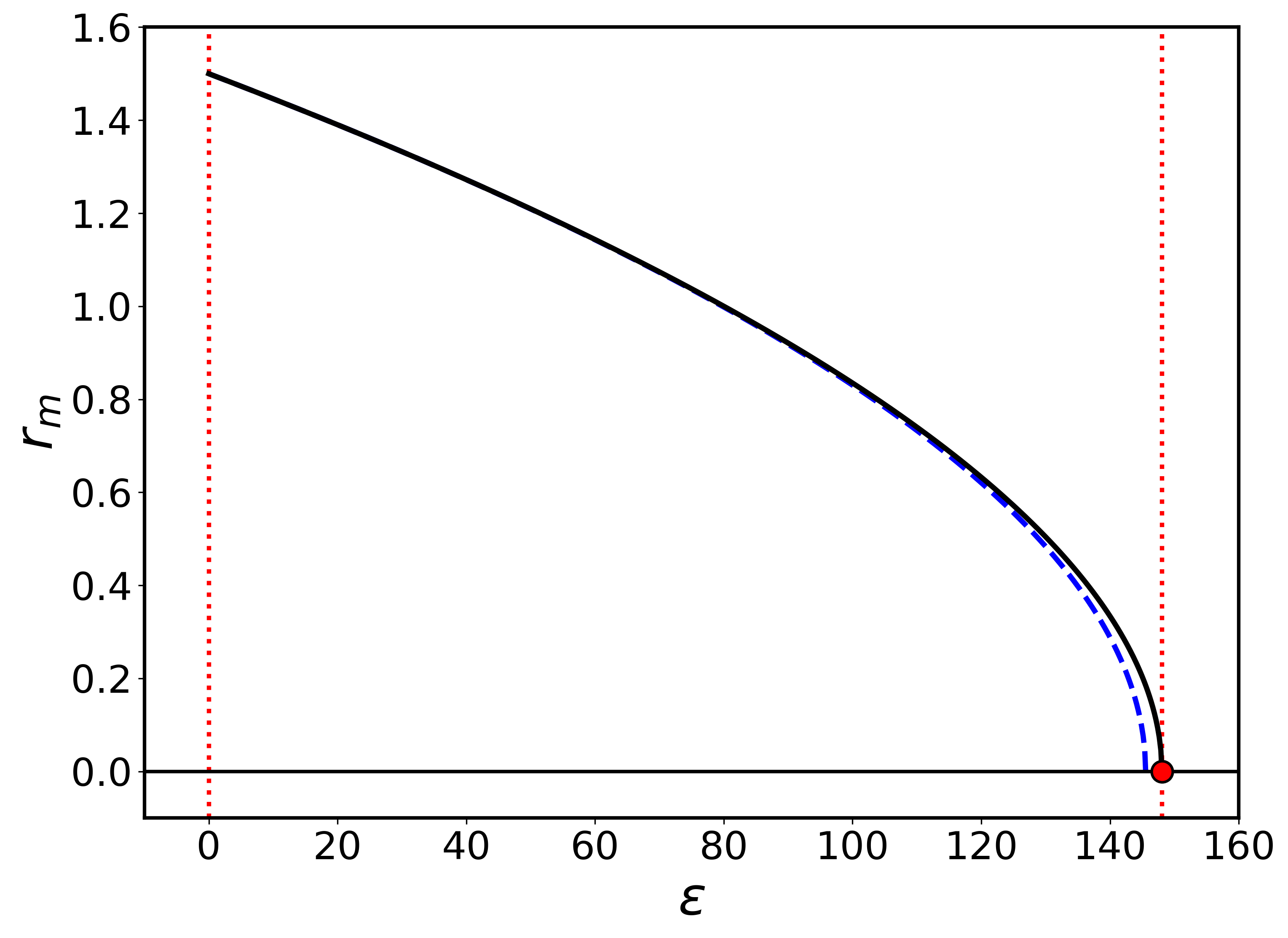}
	\end{subfigure}\hfill
	\begin{subfigure}{.49\textwidth}
		\centering
		\includegraphics[width=1\linewidth]{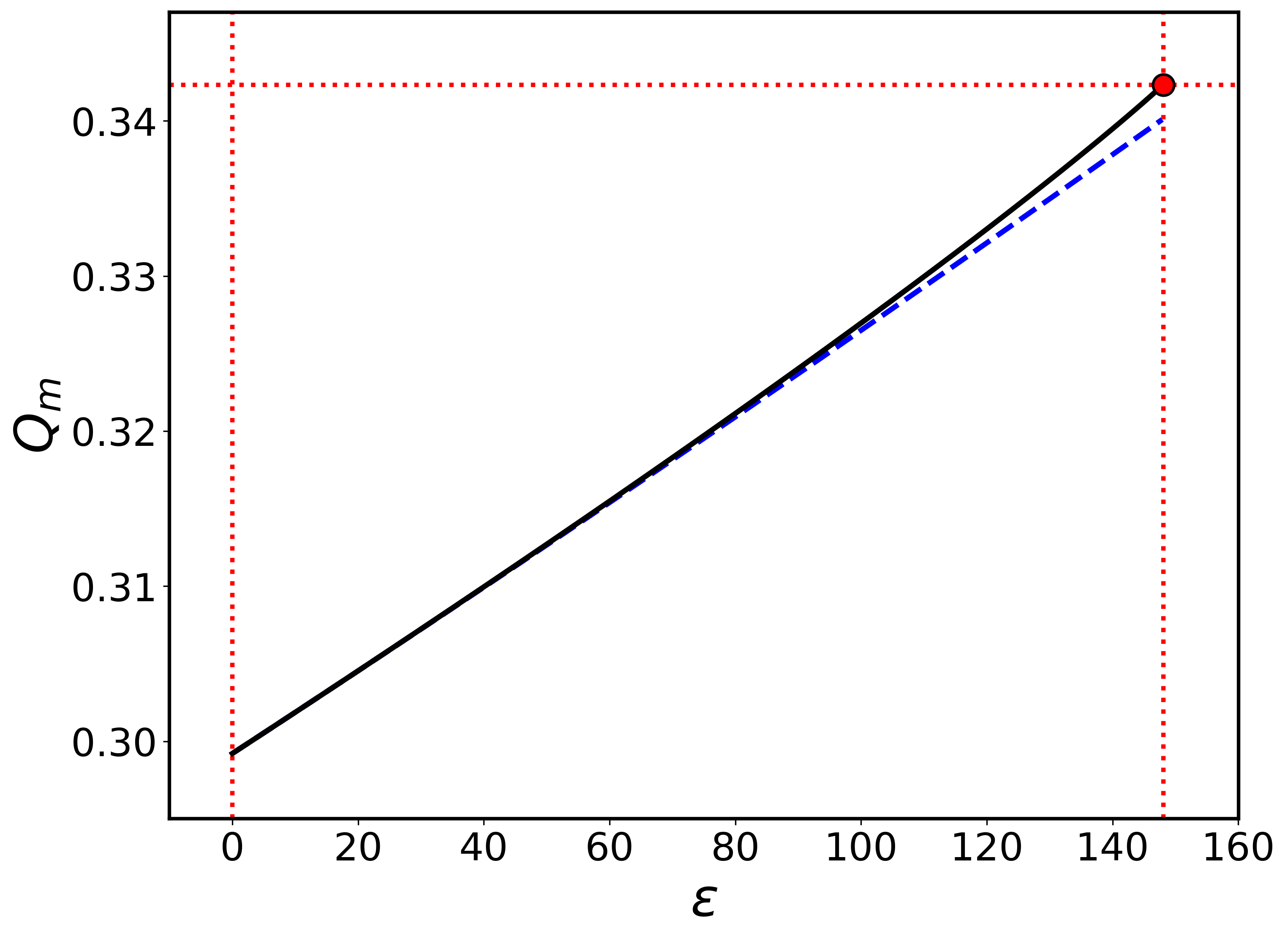} 
	\end{subfigure}
	\caption{The merger circular radius $r_{\rm m}$ (left) and merger charge $Q_{\rm m}$ (right) as functions of the parameter $\epsilon$. The solid black curves denote the full numerical solutions obtained from Eqs.~\eqref{eq:rc_crit} and~\eqref{eq:Fc_crit}, while the dashed blue curves represent the associated analytical approximations given by Eqs.~\eqref{eq:exact_numeric_rm} and~\eqref{eq:Qm_extremal}. The critical bifurcation point is highlighted by the red circular marker.}
	\label{fig:extrm_charge_radius}
\end{figure}

	Figure~\ref{fig:extrm_charge_radius} compares the numerical solutions with the analytical approximations. The numerical solutions are obtained from the simultaneous solution of Eqs.~\eqref{eq:rc_crit} and~\eqref{eq:Fc_crit}, while the analytical charge is given by Eq.~\eqref{eq:Qm_extremal} and the hybrid radius by Eq.~\eqref{eq:exact_numeric_rm}. The perturbative expression for $Q_{\rm m}$ remains very close to the numerical critical charge throughout the physical branch. The hybrid expression for $r_{\rm m}$, constructed from the same local expansion of $Q_{\rm m}$, reproduces the characteristic nonlinear decrease of the merger radius. For $\epsilon=0$, $M=1$, and $\kappa=8\pi$, Eq.~\eqref{eq:Qm_extremal} gives $Q_{\rm m}=Q_{0}=\frac{3}{2\sqrt{\kappa}}\approx 0.2992067$, while Eq.~\eqref{eq:exact_numeric_rm} gives $r_{\rm m}=R_{0}=1.5$, in exact agreement with the RN values. Thus, at $\epsilon=0$, the perturbative result is exact. As $\epsilon$ increases, the merger charge varies smoothly, while the merger radius decreases increasingly rapidly toward the bifurcation point. The discrepancy between the perturbative and numerical results consequently increases with $\epsilon$ and reaches its maximum at the bifurcation point, which makes this point particularly useful for assessing the accuracy of the expansion. To quantify this accuracy, we compare the perturbative charge with the exact bifurcation parameters obtained from Eqs.~\eqref{eq:QM_cr} and ~\eqref{eq:eps_M_cr}. The exact bifurcation charge is $Q_{\rm b} = C_{0} M$, where $
	C_{0} = \frac{4.5}{\Gamma(1/4)^2}
	\approx 0.3423342594$, corresponding to the critical parameter $\epsilon_{\rm b} = 2\kappa^2 Q_{\rm b}^2
	=2\kappa^2 C_{0}^{2}M^{2}
	\approx 148.0506923\,M^2$. Evaluating the second-order expansion~\eqref{eq:Qm_extremal} at $\epsilon_{\rm b}$ gives $Q_{\rm m}(\epsilon_{\rm b})=D_{0}M$, where
	$
	D_{0}
	=
	\sqrt{
		\frac{9}{4\kappa}
		+\frac{C_{0}^{2}}{5}
		+\frac{16\kappa C_{0}^{4}}{2025}
	}
	\approx 0.3401330998.
	$
	The corresponding absolute deviation is therefore
	$
	|Q_{\rm m}(\epsilon_{\rm b})-Q_{\rm b}|
	=
	|D_0-C_0|M
	\approx 2.20116\times10^{-3}M,
	$
	which gives the relative error
	\begin{align}\label{eq:relative_error}
		\frac{|Q_{\rm m}(\epsilon_{\rm b})-Q_{\rm b}|}{Q_{\rm b}}\times100
		=
		\frac{|D_0-C_0|}{C_0}\times100
		\approx 0.643\%.
	\end{align}
	Thus, even at the endpoint of the physical branch, the second-order perturbative expression reproduces the exact bifurcation charge with sub-$1\%$ accuracy. This demonstrates that the local expansion of $Q_{\rm m}$ captures not only the immediate neighborhood of the RN limit but also the global evolution of the critical charge to a good approximation.
	
	The situation is qualitatively different for the merger radius. At the exact bifurcation point, the numerical solution satisfies $r_{\rm m}(\epsilon_{\rm b})=0$, whereas the hybrid approximation~\eqref{eq:exact_numeric_rm} reaches zero at
	$
	\epsilon_{\rm b}^{\rm hyb}=5.79058443\,\kappa M^2.
	$
	For $\kappa=8\pi$ and $M=1$, this gives $\epsilon_{\rm b}^{\rm hyb}\approx145.53326$, to be compared with the exact value $\epsilon_{\rm b}\approx148.05069$. Hence, the hybrid approximation predicts the endpoint slightly before the exact numerical bifurcation. Nevertheless, as illustrated in Fig.~\ref{fig:extrm_charge_radius}, it captures the rapid suppression of $r_{\rm m}$ near the endpoint. The remaining discrepancy is expected, since the perturbative expansion of $Q_{\rm m}$ is organized around $\epsilon=0$, whereas the bifurcation point lies in the strongly nonlinear regime.

	The combined analytical and numerical results therefore provide a coherent description of the critical circular-orbit boundary. The perturbative expansion~\eqref{eq:Qm_extremal} accurately tracks the merger charge, while Eq.~\eqref{eq:exact_numeric_rm} translates this information into a reliable approximation for the merger radius without artificially imposing a polynomial behavior on a quantity that vanishes at the bifurcation point. The critical curve $Q_{\rm m}(\epsilon;M)$ therefore exists continuously from the RN limit up to the bifurcation point $(Q_{\rm b},\epsilon_{\rm b})$, where the marginally stable circular orbit terminates at $r=0$. This behavior is consistent with studies of circular geodesics in regular black-hole and no-horizon spacetimes, where the photon-sphere structure and characteristic orbit radii are known to depend sensitively on the charge parameter and can exhibit qualitative changes at critical values~\cite{Stuchlik2015,Chiba2017,Guo2023}.

	\subsection{Background Effective Potential: Horizonless Configurations and Black Points}\label{subsec:Veff_bg}

\begin{figure}[tb]
		\centering
		\begin{subfigure}{.5\textwidth}
			\centering
			\includegraphics[width=1\linewidth]{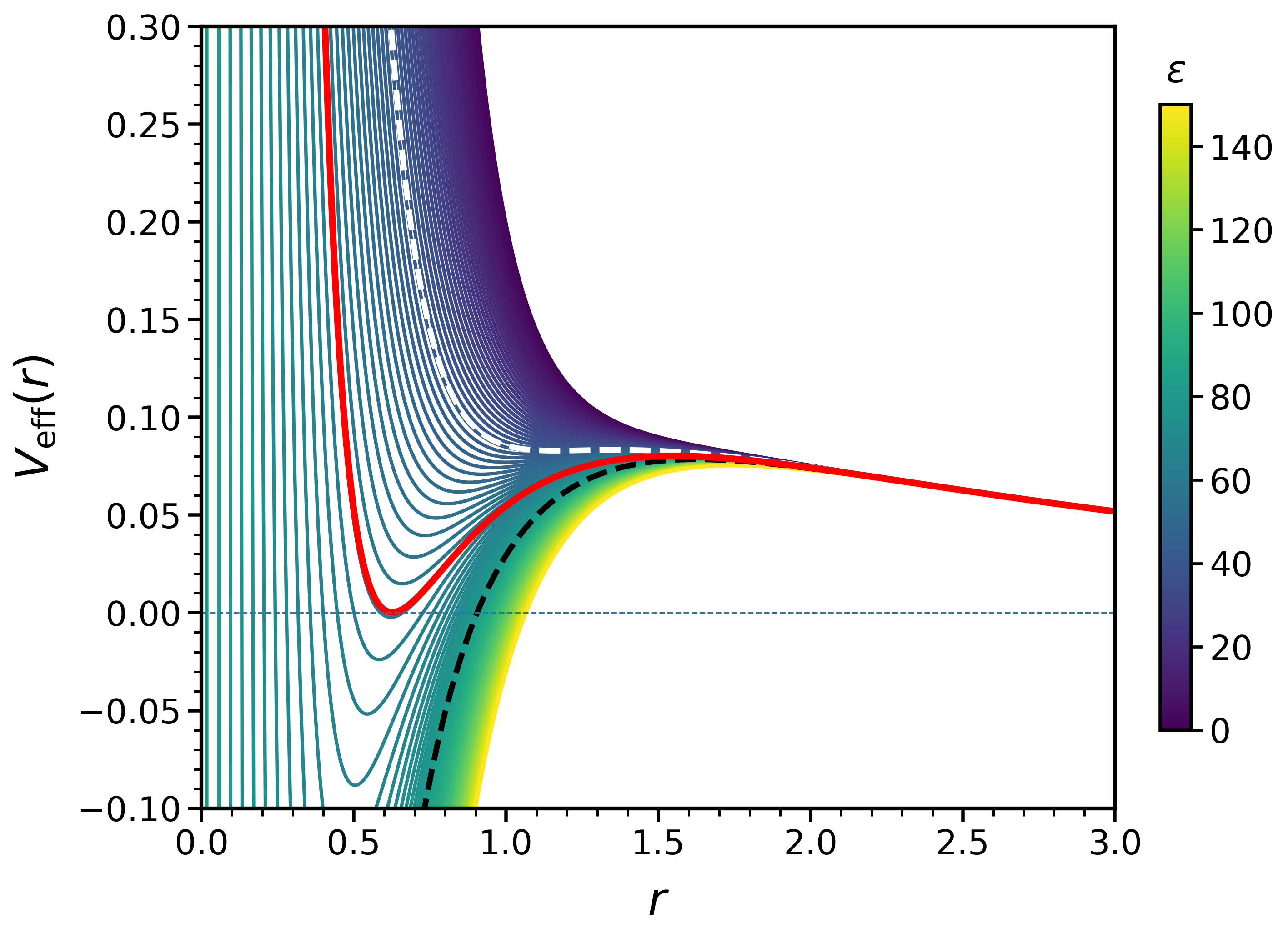}
		\end{subfigure}%
		\begin{subfigure}{.5\textwidth}
			\centering
			\includegraphics[width=1\linewidth]{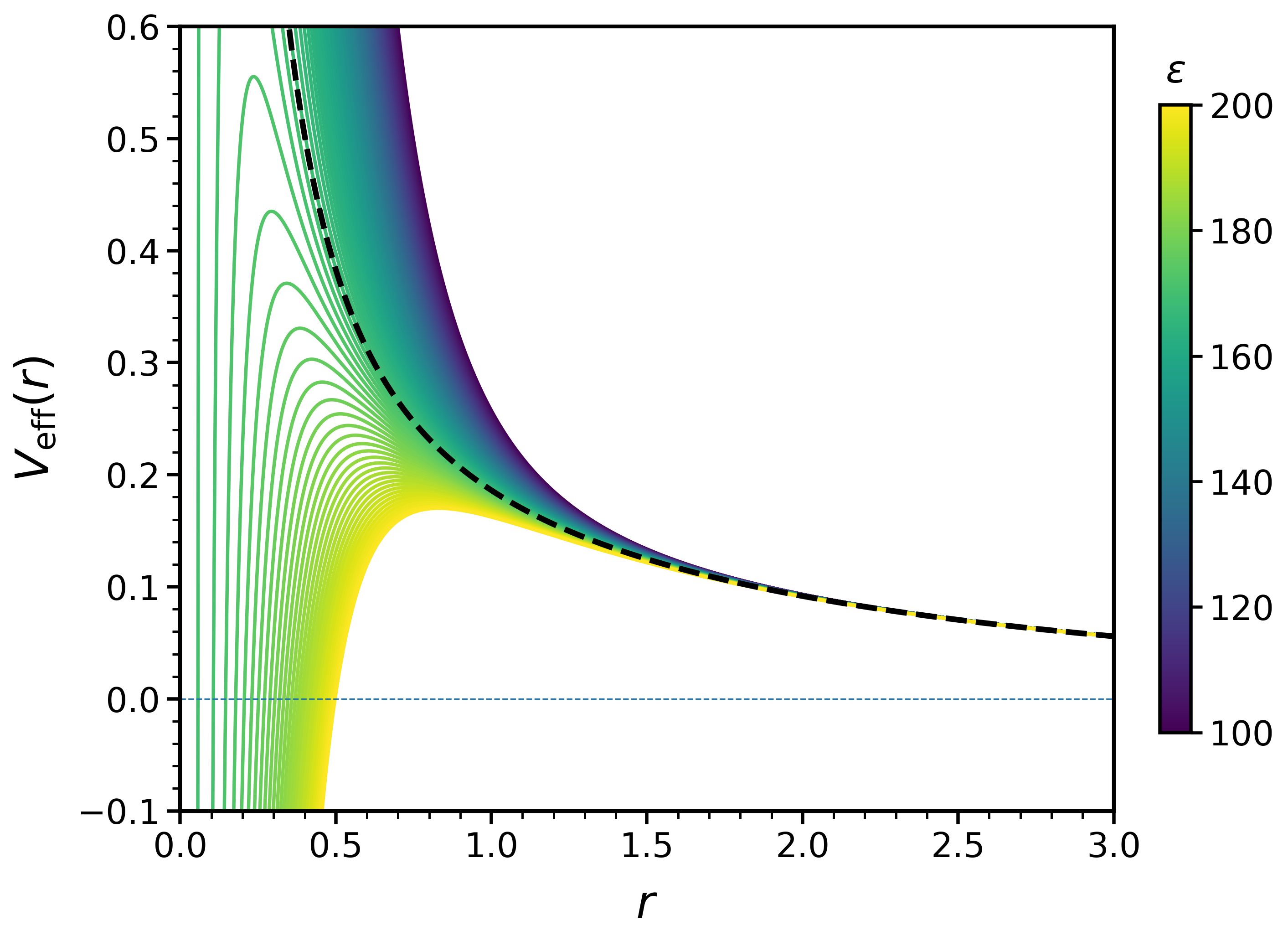}
		\end{subfigure}
		\caption{The background effective potential $V_{\rm eff}^{\rm bg}(r)$ for various values of $\epsilon$ and $Q$, with $M=1.0$, $\kappa=8\pi$, and $\Lambda=0$. The left and right panels correspond to $Q=0.31$ and $Q=0.35$, respectively. The dashed white curve (inflection point) corresponds to the critical boundary separating configurations with no extrema from those exhibiting both a local minimum and a local maximum. The red solid curve represents a point on the extremal horizon boundary, while the dashed black curve marks a point on the- non-critical boundary (zero-radius orbit). The latter separates configurations with only a local maximum from those either exhibiting both a local minimum and a local maximum (left panel) or lacking extrema altogether (right panel).}
		\label{fig:Veff_bg}
	\end{figure}

	We now examine the background effective potential $V_{\rm eff}^{\rm bg}(r)$ from Eq.~\eqref{eq:Veff_bg_r} for different values of $\epsilon$ at fixed $Q$. The result is shown in Fig.~\ref{fig:Veff_bg}. As anticipated from the stability analysis, the inner orbit corresponds to a local minimum and the outer orbit to a local maximum, so that the inner orbit is stable and the outer orbit is unstable. On the critical boundary $Q_{\rm m}(\epsilon;M)$, the potential develops an inflection point where the two orbits merge into a degenerate marginally stable orbit. This is indicated by the dashed white curve in the left panel, which corresponds to $Q=0.31<Q_{\rm b}$ and $\epsilon\approx 41.5<\epsilon_{\rm b}$. The solid red curve marks a point on the extremal horizon boundary $Q_{\rm ext}(\epsilon;M)$ at $\epsilon\approx 61.3$.
	
	On the zero-radius non-critical boundary, where $C(\epsilon,Q;M)=0$, i.e., $Q=Q_{\rm nc}(\epsilon;M)=\gamma M^{2/3}\epsilon^{1/6}$, the effective potential develops either an infinitely deep minimum or an infinitely high barrier, depending on whether $\epsilon<\epsilon_{\rm b}$ or $\epsilon\ge\epsilon_{\rm b}$. To see this, we take the limit $r\to0$ with $C(\epsilon,Q_{\rm nc};M)=0$ and obtain
	\begin{align}\label{eq:Veff_bg_limit}
		\lim_{r\to0} V_{\rm eff}^{\rm bg}(r)
		\approx	\frac{L^{2}}{r^{2}}
		\left(
		1-\frac{\sqrt{2}\kappa |Q_{\rm nc}|}{\sqrt{\epsilon}}
		-\frac{\kappa r^{2}}{3\epsilon}
		\ln\left[
		\sqrt{\frac{2}{\epsilon}}\frac{r^{2}}{|Q_{\rm nc}|}
		\right]
		+ \cdots
		\right).
	\end{align}
	
	For $\epsilon < 2\kappa^2 Q_{\rm nc}^2$, the outer horizon and null circular orbit have finite radii, while the corresponding radii of the inner branches shrink to zero on the $Q_{\rm nc}$ boundary. Here, the potential minimum diverges as $V_{\rm eff}^{\rm bg}(r\to0)\propto -1/r^{2}\to-\infty$. This is indicated by the dashed black curve in the left panel, corresponding to $Q_{\rm nc}=\gamma M^{2/3}\epsilon^{1/6}=0.31<Q_{\rm b}$, which yields $\epsilon\approx 81.6<\epsilon_{\rm b}$. Since the Kretschmann scalar diverges at $r=0$~\cite{Soleng1995}, the potential minimum is not a trap. Geodesics that pass over the finite barrier at $r_{\max}>0$ cross the event horizon and terminate at the singularity. 

	For $\epsilon > 2\kappa^{2}Q_{\rm nc}^{2}$, we take $Q_{\rm nc} = \gamma M^{2/3}\epsilon^{1/6} = 0.35$, which exceeds $Q_{\rm b}$ and gives $\epsilon \approx 169.5 > \epsilon_{\rm b}$. In this regime the inner branches disappear, and both the outer horizon and the outer null circular orbit shrink to zero as one approaches the $Q_{\rm nc}$ boundary, leaving behind an infinitely small, point-like structure known as an ordinary black point~\cite{Sokolov2025}. But because $f(0) \neq 0$, the $Q_{\rm nc}$ boundary where $r_{\rm h} = 0$ actually corresponds to a horizonless configuration. For both the black point and this horizonless configuration, $V_{\rm eff}^{\rm bg}(r\to0)\propto 1/r^{2}\to+\infty$, so the potential forms an infinite barrier with no extrema, and only radial geodesics can reach $r=0$. This curve is shown as the dashed black line in the right panel.
	
	Furthermore, at the bifurcation point, all horizons and null circular orbits shrink to $r=0$, with $Q_{\rm nc}=Q_{\rm b}$ corresponding to $\epsilon=\epsilon_{\rm b}=2\kappa^{2}Q_{\rm nc}^{2}$. This yields $f(r_{\rm h}^{\rm ext}=0)=0$, implying that $r=0$ is a genuine horizon solution and leaving a point-like configuration known as extremal black point~\cite{Soleng1995}. Here the logarithmic term dominates and produces an infinite barrier, $V_{\rm eff}^{\rm bg}(0)\propto-\ln(r)\to+\infty$, which does not exhibit extrema. 
	
	Thus, for the present LNED model, both types of black point and the horizonless configuration at the $Q_{\rm nc}$ boundary correspond to a purely gravitational repulsive core that prevents non-radial background null geodesics from reaching the singularity at $r=0$. This effect is purely gravitational: as we show in Subsec.~\ref{subsec:Veff_ph}, NED photons follow the effective geometry and are not prevented from reaching the singularity.
		
	The analysis above also reveals another interesting class of horizonless configurations, which we now examine. For $Q=0.31$, a horizonless region appears between the marginal curve (dashed white) and the extremal curve (solid red). In this region the effective potential remains non-negative and diverges as $r\to0$, resulting in an infinite central barrier. Since $C<0$ (i.e., $Q>Q_{\rm nc}$ in this region), we obtain for $r\to0$
	\begin{align}\label{eq:Veff_bg_limit_2}
		\lim_{r\to0} V_{\rm eff}^{\rm bg}(r)
		\approx	\frac{L^{2}}{r^{2}}
		\left(
		- \frac{2C(\epsilon,Q;M)}{r}
		+ 1 
		-\frac{\sqrt{2}\kappa |Q_{\rm nc}|}{\sqrt{\epsilon}}
		-\frac{\kappa r^{2}}{3\epsilon}
		\ln\left[
		\sqrt{\frac{2}{\epsilon}}\frac{r^{2}}{|Q_{\rm nc}|}
		\right]
		+ \cdots
		\right),
	\end{align}
	which gives $V_{\rm eff}^{\rm bg}(r\to0)\propto 1/r^{3}\to+\infty$. This barrier hides the curvature singularity at $r=0$ from non-radial geodesics. In contrast to the repulsive core discussed above, here both stable and unstable circular orbits exist at finite radii. The stable minimum can confine massless test particles within a finite radial region for energies in the range $V_{\rm min}<E^{2}<V_{\rm max}$, while the unstable local maximum marks the boundary between trapped and escaping trajectories. Similar bound orbit structures have been studied in regular black holes and horizonless configurations~\cite{Stuchlik2015,Chiba2017,Habibina2021,Toshmatov2026}. This trapping is classical and metastable: the barrier at $r_{\rm max}$ is finite, so perturbations or quantum tunneling can eject particles from the well. The stability of the circular orbits themselves can be characterized by the Lyapunov exponent, which in the eikonal limit is related to the quasinormal mode frequencies~\cite{Mondal2020}. We emphasize again that this analysis concerns massless test particle motion in the background geometry and should not be interpreted as a description of photon propagation in the nonlinear electromagnetic sector.

	Having established the structure of circular null geodesics of the background spacetime, we next distinguish these geometrical null orbits from the physical propagation of electromagnetic waves in LNED. Because nonlinear electrodynamics modifies the characteristic surfaces of electromagnetic perturbations, photons generally propagate along null geodesics of an effective optical metric rather than those of the background metric. We therefore turn next to the effective optical geometry and determine the corresponding physical photon-sphere structure.

	\section{Photon Motion in the Effective Optical Geometry}\label{sec:PMEG}
	
	The circular null geodesics discussed in the preceding section characterize the propagation of massless test particles in the background spacetime described by Eq.~\eqref{eq:metric}. For nonlinear electrodynamics, however, these background null geodesics do not, in general, coincide with the trajectories followed by electromagnetic waves. The nonlinear electromagnetic interaction modifies the characteristic surfaces of high-frequency electromagnetic perturbations, so that the physical photon trajectories are determined by an effective optical geometry rather than directly by the background metric~\cite{Novello2000,DeLorenci2000}. The same nonlinear electromagnetic sector also leaves signatures in the perturbation spectrum of the background, as shown for spacetimes in general relativity coupled to nonlinear electrodynamics~\cite{Toshmatov2019}, and the causality of the resulting effective characteristics has been analyzed in detail in~\cite{dePaula2024}. The purpose of this section is therefore to construct the corresponding effective metric for the LNED solution obtained in Secs.~\ref{sec:action_integral}--\ref{sec:field_equations} and to determine the associated physical photon sphere.
	
	For a nonlinear theory whose Lagrangian depends only on the electromagnetic invariant \(\mathcal F\), the characteristic equation for electromagnetic perturbations can be written in terms of an effective inverse metric as~\cite{Novello2000,DeLorenci2000}
	\begin{align}\label{eq:effective_inverse_metric}
		\tilde g^{\mu\nu}k_\mu k_\nu
		&= 0, 	
		\nonumber \\
		\tilde g^{\mu\nu}
		&=
		g^{\mu\nu}
		-4\frac{\mathcal L_{\mathcal F\mathcal F}}{\mathcal L_{\mathcal F}}
		F^\mu{}_{\alpha}F^{\alpha\nu}.
	\end{align}
	The same effective-metric structure underlies the analysis of birefringence and high-frequency modes in other nonlinear electromagnetic theories, as illustrated for the Einstein--Euler--Heisenberg black hole~\cite{Breton2021}. For the LNED model introduced in Eq.~\eqref{eq:L_electro}, the required first and second derivatives are 
	\begin{align}\label{eq:L_derivatives}
		\mathcal L_{\mathcal F} 
		&= 
		-\frac{1}{4\left(1+\epsilon\mathcal F/4\right)} 
		\nonumber \\
		\mathcal L_{\mathcal F\mathcal F}
		&= \frac{\epsilon}
		{16\left(1+\epsilon\mathcal F/4\right)^2},
	\end{align} 
	and hence
	\begin{align}\label{eq:effective_metric_coeff}
		\tilde g^{\mu\nu}
		=
		g^{\mu\nu}
		+
		\frac{\epsilon}
		{1+\epsilon\mathcal F/4}
		F^\mu{}_{\alpha}F^{\alpha\nu}.
	\end{align}
	Using Eqs.~\eqref{eq:invariant} and~\eqref{eq:charge_cons}, the factor \(1+\epsilon\mathcal F/4\) can be expressed through \(\Delta(r)\), which motivates the definition below. Using the purely electric configuration and the electric field already obtained in Eq.~\eqref{eq:electric_field}, it is convenient to introduce
	\begin{equation}\label{eq:Delta}
		\Delta(r)\equiv\sqrt{r^4+2\epsilon Q^2}.
	\end{equation}
	Substitution into the effective inverse metric in Eq.~\eqref{eq:effective_metric_coeff} then gives 
	$\tilde g^{tt} = -\frac{\Delta}{r^2 f}$, 
	$\tilde g^{rr}=\frac{\Delta}{r^2}f$, 
	$\tilde g^{\theta\theta}=\frac1{r^2}$, 
	and 
	$\tilde g^{\varphi\varphi}
	= \frac1{r^2\sin^2\theta}$. 
	Since the effective inverse metric is diagonal in $(t,r,\theta,\varphi)$, the inversion is straightforward. The effective line element can therefore be written as
	$
	d\tilde s^2
	=
	-\frac{r^2f(r)}{\Delta(r)}\,dt^2
	+\frac{r^2}{\Delta(r)f(r)}\,dr^2
	+r^2d\Omega^2.
	$
	Since null geodesics are invariant under conformal transformations, the common factor \(r^2/\Delta\) can be removed. The effective optical geometry is consequently represented by
	\begin{align}\label{eq:effective_line_element}
		d\tilde s^2
		=
		-f(r)\,dt^2
		+\frac{dr^2}{f(r)}
		+\Delta(r)
		\left(d\theta^2+\sin^2\theta\,d\varphi^2\right).
	\end{align}
	Thus, relative to the background metric~\eqref{eq:metric}, the essential modification induced by LNED is the replacement of the areal factor \(r^2\) in the angular sector by \(\Delta(r)\). This distinction is precisely what separates the physical photon sphere from the background circular null orbits obtained in the preceding section.

	\subsection{Photon Sphere and Photon Circular Orbits} \label{subsec:photon_sphere_orbits}
	The geodesic analysis of the effective metric proceeds in complete analogy with that of Sec.~\ref{sec:PMBG}. Restricting to the equatorial plane, the Killing symmetries associated with \(t\) and \(\varphi\) give
	\begin{align}\label{eq:photon_conserved}
		E &= f(r)\dot t,
		\nonumber \\
		L &= \Delta(r)\dot\varphi,
	\end{align}
	where \(E\) and \(L\) denote the conserved energy and angular momentum of the effective null trajectory. The null condition then yields
	\begin{align}\label{eq:photon_null_condition}
		\dot r^2
		+
		V_{\rm eff}^{\rm ph}(r)
		=
		E^2,	
	\end{align}
	where the effective radial potential for physical photons is
	\begin{align}\label{eq:Veff_ph_r}
		V_{\rm eff}^{\rm ph}(r)
		=
		\frac{L^2f(r)}{\Delta(r)}.	
	\end{align}
	A circular photon orbit is therefore determined by the simultaneous conditions
	\begin{align}\label{eq:photon_circular_conditions}
		V_{\rm eff}^{\rm ph}(r_{\rm ph})
		=
		E^2,
	\end{align}
	and
	\begin{align}\label{eq:photon_circular_derivative}
		\left.
		\frac{dV_{\rm eff}^{\rm ph}}{dr}
		\right|_{r=r_{\rm ph}}=0.
	\end{align}
	Since $\frac{d\Delta(r)}{dr}=\frac{2r^3}{\Delta(r)}$, the photon-sphere condition becomes
	\begin{equation}\label{eq:photon_sphere_condition}
		\left(r_{\rm ph}^4+2\epsilon Q^2\right)
		\left.
		\frac{df(r)}{dr}
		\right|_{r=r_{\rm ph}}
		-
		2r_{\rm ph}^3 f(r_{\rm ph})
		=0 .
	\end{equation}
	For the present LNED solution, Eq.~\eqref{eq:photon_sphere_condition} can be evaluated without re-differentiating the full expression for \(f(r)\) in Eq.~\eqref{eq:exact_fr}. In particular, the first-order metric differential equation~\eqref{eq:diff_f} gives
	$
	r_{\rm ph} f'(r_{\rm ph})+f(r_{\rm ph})
	=
	1-\Lambda r_{\rm ph}^2-\kappa r_{\rm ph}^2\rho(r_{\rm ph}),
	$
	where the energy density \(\rho(r_{\rm ph})\) is already given by Eq.~\eqref{eq:rho}. Using this relation in Eq.~\eqref{eq:photon_sphere_condition} leads directly to the compact form
	\begin{align}\label{eq:photon_sphere_reduced}
		\left(r_{\rm ph}^4+2\epsilon Q^2\right)
		\left[
		1-\Lambda r_{\rm ph}^2-\kappa r_{\rm ph}^2\rho(r_{\rm ph})
		\right]
		-
		\left(3r_{\rm ph}^4+2\epsilon Q^2\right)f(r_{\rm ph})
		=0.
	\end{align}
	Substituting the energy density \(\rho(r_{\rm ph})\) from Eq.~\eqref{eq:rho} and the metric function \(f(r_{\rm ph})\) from Eq.~\eqref{eq:exact_fr} into Eq.~\eqref{eq:photon_sphere_reduced} yields a single explicit equation for the physical photon-sphere radius. After collecting the different contributions, it takes the form
	\begin{align}\label{eq:photon_master}
		F(r_{\rm ph}, \epsilon, Q; M)
		={}&
		\left(3r_{\rm ph}^{4}+2\epsilon Q^{2}\right)
		C(\epsilon,Q;M)
		-\frac{1}{3}\Lambda\epsilon Q^{2}r_{\rm ph}^{3}
		\nonumber\\
		&-\frac{\kappa Q^{2}r_{\rm ph}^{3}}{3}
		\ln\left(
		\frac{
			r_{\rm ph}^{2}\sqrt{r_{\rm ph}^{4}+2\epsilon Q^{2}}
			-r_{\rm ph}^{4}}
		{\epsilon Q^{2}}
		\right)
		\nonumber\\
		&+\frac{\kappa \, r_{\rm ph}}{9\epsilon}
		\left(3r_{\rm ph}^{4}-4\epsilon Q^{2}\right)
		\sqrt{r_{\rm ph}^{4}+2\epsilon Q^{2}}
		+\frac{4\kappa Q^{2}r_{\rm ph}^{3}}{9}
		- r_{\rm ph}^{5}
		-\frac{\kappa r_{\rm ph}^{7}}{3\epsilon}
		\nonumber\\
		&+\frac{2\kappa |Q| r_{\rm ph}}{9}
		\sqrt{\frac{2}{\epsilon}}
		\left(3 r_{\rm ph}^{4}+2\epsilon Q^{2}\right)
		{}_2F_1\left(
		\frac14,\frac12;\frac54;
		-\frac{r_{\rm ph}^{4}}{2\epsilon Q^{2}}
		\right) = 0.
	\end{align}
	
	Equation~\eqref{eq:photon_master} is the master equation for the physical photon-sphere radius in the LNED effective geometry. Its structure makes explicit the distinction between the background circular-null problem considered in Sec.~\ref{sec:PMBG} and the actual propagation of electromagnetic perturbations. Both problems depend on the same metric function $f(r)$, and therefore on the same gravitational backreaction of the nonlinear electromagnetic field, but the physical photon trajectory is additionally governed by the optical factor $\Delta(r)$, which replaces the background angular factor $r^2$. This distinction is central to the physics: photons propagate along null trajectories of an effective metric determined by the electromagnetic background rather than along the null geodesics of the background spacetime~\cite{Novello2000,DeLorenci2000,Kruglov2019,Habibina2021,Tang2023}. Nonlinear electrodynamics therefore modifies photon propagation through both the spacetime geometry and the effective optical geometry, which is the central reason why the photon-sphere structure need not coincide with that obtained from the background metric alone.
	\begin{figure}[t!]
		\centering
		\begin{subfigure}{0.49\textwidth}
			\centering
			\includegraphics[width=\linewidth]{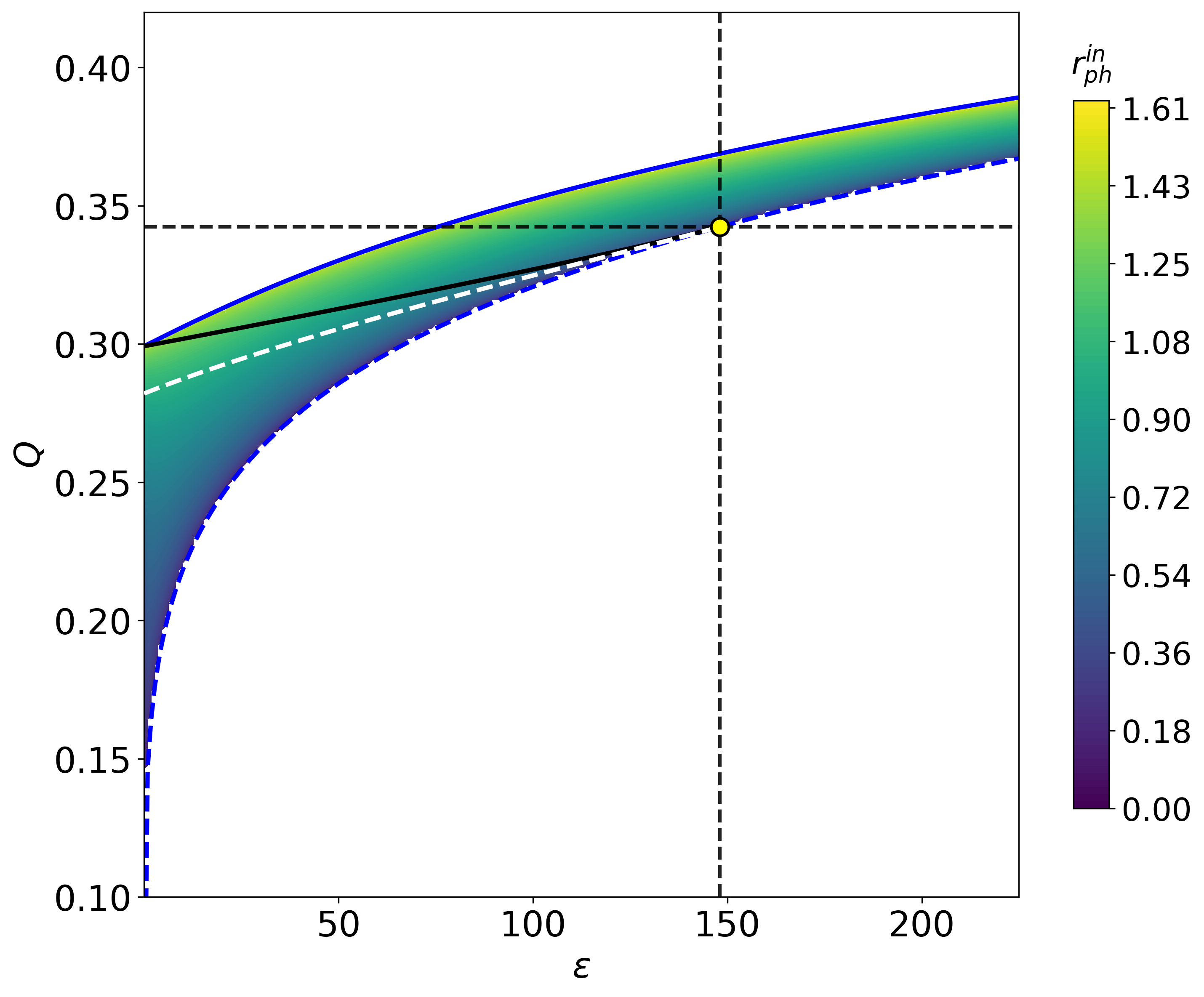}
			\caption{\footnotesize{Inner stable photon orbit}}
		\end{subfigure}
		\hfill
		\begin{subfigure}{0.49\textwidth}
			\centering
			\includegraphics[width=\linewidth]{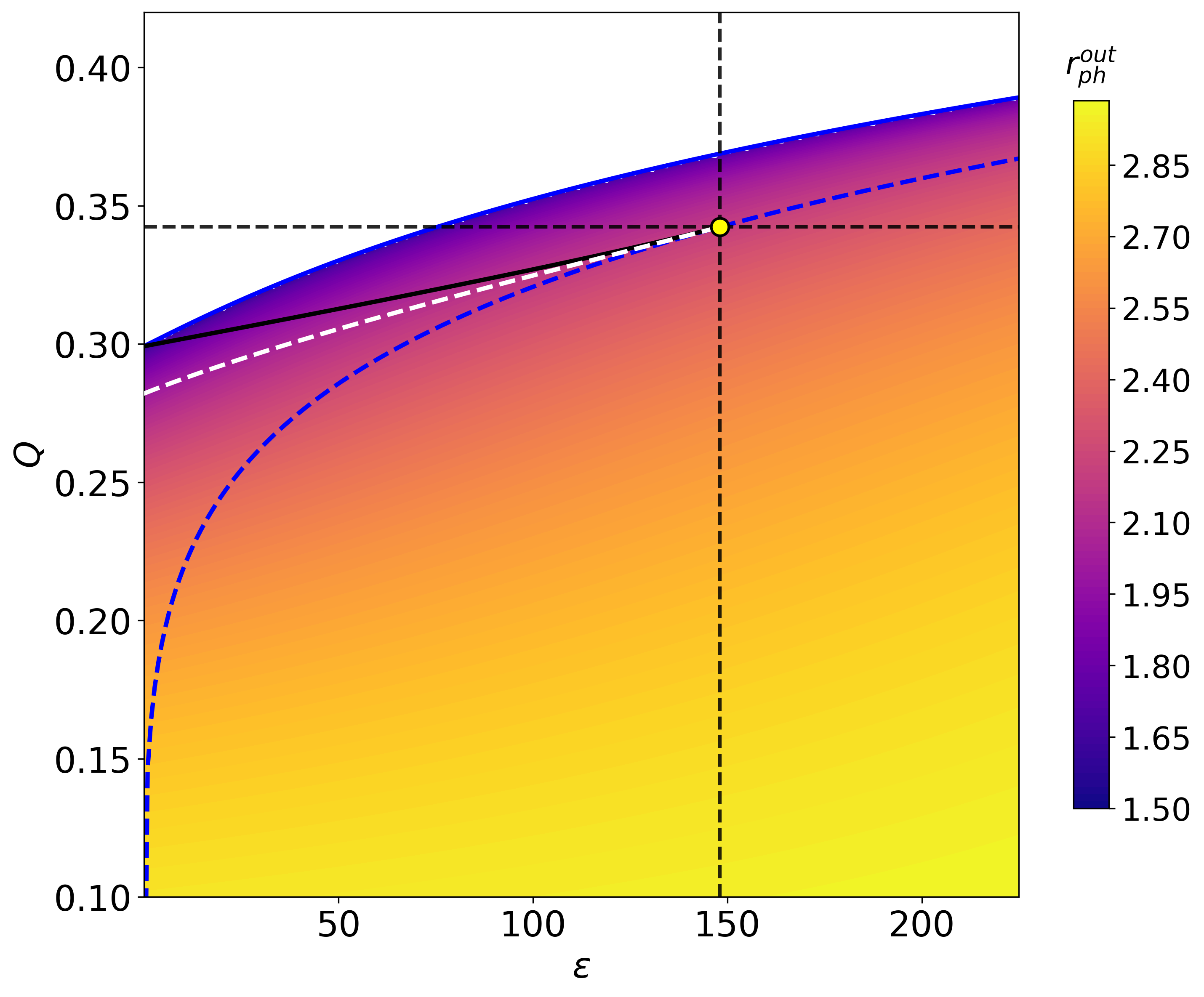}
			\caption{\footnotesize{Outer unstable photon sphere}}
		\end{subfigure}
		\caption{Effective photon-sphere radii in the $(Q,\epsilon)$ parameter space for $M=1.0$, $\kappa=8\pi$, and $\Lambda=0$. The left and right panels show the inner stable photon orbit and outer unstable photon sphere, respectively. The solid black curve denotes the critical merger boundary of the background circular-null orbits, while the extremal horizon boundary is represented by a dashed white curve. In addition, the solid and dashed blue curves denote the numerical critical merger and analytical non-critical boundaries of the effective photon-sphere structure, respectively. The yellow circle marks the horizon/background zero-radius bifurcation point $(Q_{\rm b},\epsilon_{\rm b})$. It is shown for reference and is not an endpoint of the effective photon-sphere merger curve.}
		\label{fig:photon_sphere}
	\end{figure}

	The numerical solutions of Eq.~\eqref{eq:photon_master} reveal two branches of circular photon trajectories, as shown in Fig.~\ref{fig:photon_sphere}. The outer branch is the physically relevant unstable photon sphere associated with the critical scattering trajectory, while the inner branch represents an additional circular photon orbit generated by the nonlinear optical geometry. The coexistence of two branches is consistent with earlier LNED studies, where nonlinear photon propagation was found to support stable bound photon configurations~\cite{Habibina2021}; the new feature here is that the two branches are embedded in a complete $(Q,\epsilon)$ phase structure and can be followed relative to the background circular-null and horizon boundaries.
	
	Two characteristic boundaries determine this structure. The first is the non-critical boundary, obtained by taking the limit $r_{\rm ph}\rightarrow0$ directly in Eq.~\eqref{eq:photon_master}. For $\epsilon>0$ and $Q\neq0$, the leading contribution gives $C(\epsilon,Q;M)=0$, and hence
	$
	Q_{\rm nc}(\epsilon;M)=\gamma M^{2/3}\epsilon^{1/6}.
	$
	This is the same parameter boundary obtained from the zero-radius limit of the horizon equation in Sec.~\ref{sec:horizon_phase}. The coincidence is therefore not numerical but follows analytically from the common mass-normalization function $C(\epsilon,Q;M)$. Nevertheless, the physical interpretation of the boundary is different for the two problems because the background and effective geometries assign different angular structures to photon propagation.
	
	The second boundary is the critical photon-sphere merger curve. It is obtained by requiring Eq.~\eqref{eq:photon_master} to possess a double root in $r_{\rm ph}$; equivalently, Eq.~\eqref{eq:photon_master} and its first radial derivative 
	\begin{align}\label{eq:photon_master_derivative}
		\frac{\partial F}{\partial r_{\rm ph}}={}&
		12r_{\rm ph}^{3}C(\epsilon,Q;M)
		-\Lambda\epsilon Q^2r_{\rm ph}^{2}
		+\frac{2\kappa Q^2r_{\rm ph}^{2}}{3}
		-5r_{\rm ph}^{4}
		-\frac{7\kappa r_{\rm ph}^{6}}{3\epsilon}
		\nonumber\\
		&+\frac{\kappa r_{\rm ph}^{4}
			\left(7r_{\rm ph}^{4}+12\epsilon Q^2\right)}
		{3\epsilon\sqrt{r_{\rm ph}^{4}+2\epsilon Q^2}}
		-\kappa Q^2r_{\rm ph}^{2}
		\ln\!\left[
		\frac{
			r_{\rm ph}^{2}\sqrt{r_{\rm ph}^{4}+2\epsilon Q^2}
			-r_{\rm ph}^{4}}
		{\epsilon Q^2}
		\right]
		\nonumber\\
		&+\frac{8\kappa |Q|r_{\rm ph}^{4}}{3}
		\sqrt{\frac{2}{\epsilon}}\,
		{}_2F_1\!\left(
		\frac{1}{4},\frac{1}{2};
		\frac{5}{4};
		-\frac{r_{\rm ph}^{4}}{2\epsilon Q^2}
		\right)=0.
	\end{align}
	are solved simultaneously. In practice, this coupled system must be solved numerically in most cases of interest, and only in certain limiting regimes can analytical progress be made. This yields two relations,
	\begin{align}\label{eq:photon_merger_relations}
		r_{\rm m}^{\rm ph}=r_{\rm m}^{\rm ph}(\epsilon;M),
		\qquad
		Q_{\rm m}^{\rm ph}=Q_{\rm m}^{\rm ph}(\epsilon;M),
	\end{align}
	which determine the finite-radius coalescence of the inner and outer effective photon-sphere branches. In the Maxwell limit, $\epsilon=0$, these relations reduce to the Reissner--Nordstr\"om values obtained earlier in Eqs.~\eqref{eq:RN_Q0} and~\eqref{eq:RN_R0}, $Q_{0} = Q_{\rm m}^{\rm ph}(0,M) = \frac{3M}{2\sqrt{\kappa}}$ and $R_{0} = r_{\rm m}^{\rm ph}(0,M) = \frac{3M}{2}$, so that the background circular-null structure and the physical photon-sphere structure share the same RN starting point. Once nonlinear electrodynamics is switched on, however, their merger curves separate because the physical photon condition contains the additional optical factor $\Delta(r)$. The present phase diagram therefore provides information that is not contained in the horizon structure or in the background null-geodesic analysis alone.
	
	For $\epsilon\ge\epsilon_{\rm b}$, this distinction becomes especially pronounced. On the non-critical curve $Q=Q_{\rm nc}$, both the outer background circular-null branch and the inner effective photon branch reach zero radius, while the outer unstable photon sphere remains at a finite positive radius. Thus the black points on this boundary possess no finite-radius background null circular orbit, yet they do support unstable photon spheres of finite radius. Moreover, for $Q<Q_{\rm nc}$, the outer branches of the photon sphere, the horizon, and the background orbit are all present. In the interval
	$
	Q_{\rm nc}<Q<Q_{\rm m}^{\rm ph},
	$
	both effective photon-sphere branches coexist in a horizonless region, even though the background geometry admits no finite positive circular-null orbit. Hence, in the strongly nonlinear regime, photon trapping and critical photon scattering can persist even after the corresponding background structure is lost. For $Q>Q_{\rm m}^{\rm ph}$, the two effective photon branches have merged and disappeared. 
	
	In addition, for $\epsilon<\epsilon_{\rm b}$, both the background circular-null orbits and the effective photon spheres can extend into the horizonless region, but the domain of the effective photon spheres extends farther. Consequently, there are parameter ranges in which the effective geometry contains both inner and outer photon orbits while the background spacetime has no finite circular-null orbit. The principal physical conclusion is therefore that the background metric alone is insufficient to determine the strong-field propagation of electromagnetic photons in LNED: the effective optical geometry introduces an additional structure whose phase boundaries are distinct from those of the background geometry, while reducing continuously to the standard RN photon dynamics in the Maxwell limit.

	\subsection{Photon Effective Potential in Horizonless and Black-Point regimes} \label{subsec:Veff_ph}
	
\begin{figure}[t!]
		\centering
		\begin{subfigure}{0.49\textwidth}
			\centering
			\includegraphics[width=\linewidth]{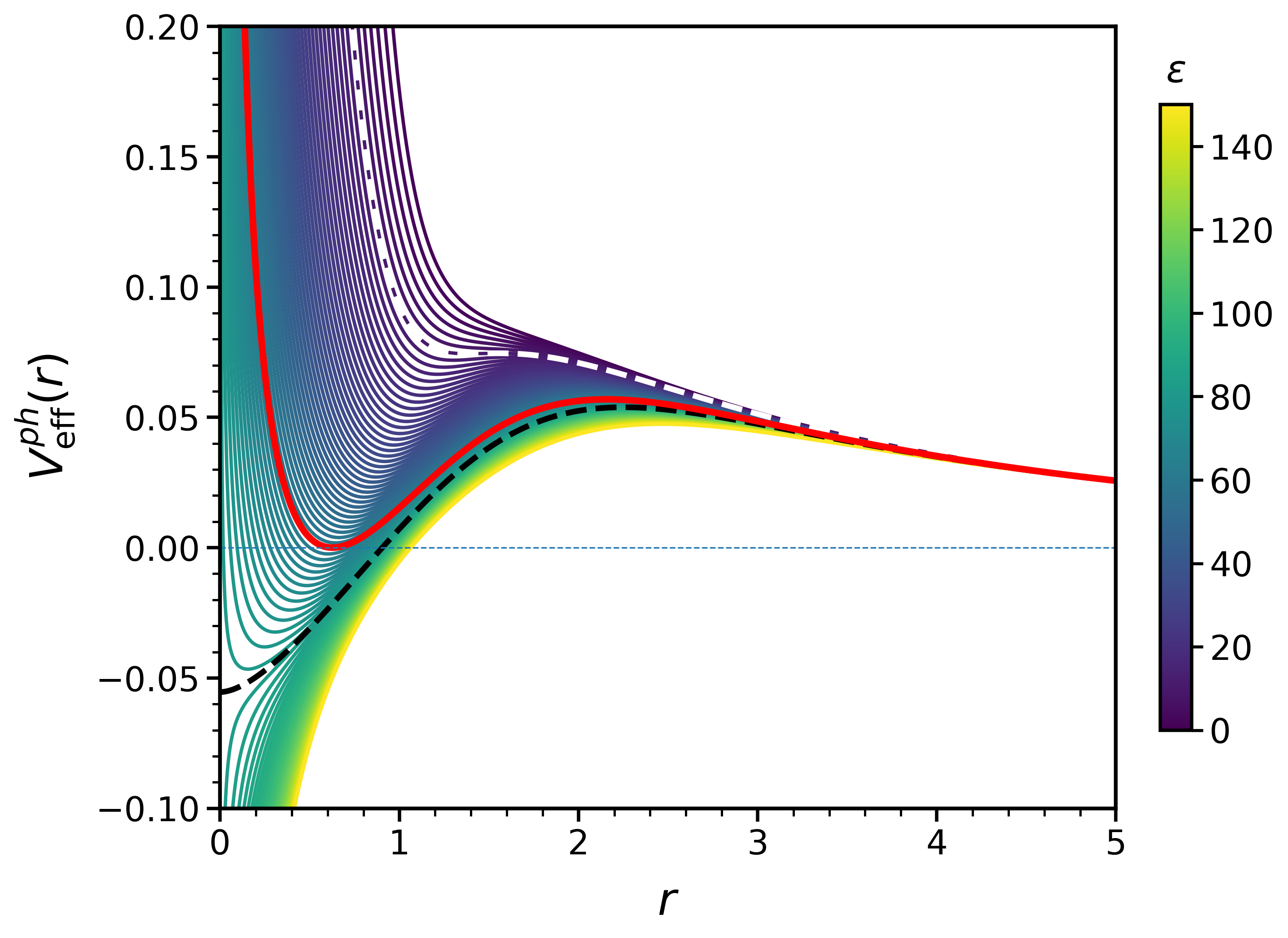}
		\end{subfigure}
		\hfill
		\begin{subfigure}{0.49\textwidth}
			\centering
			\includegraphics[width=\linewidth]{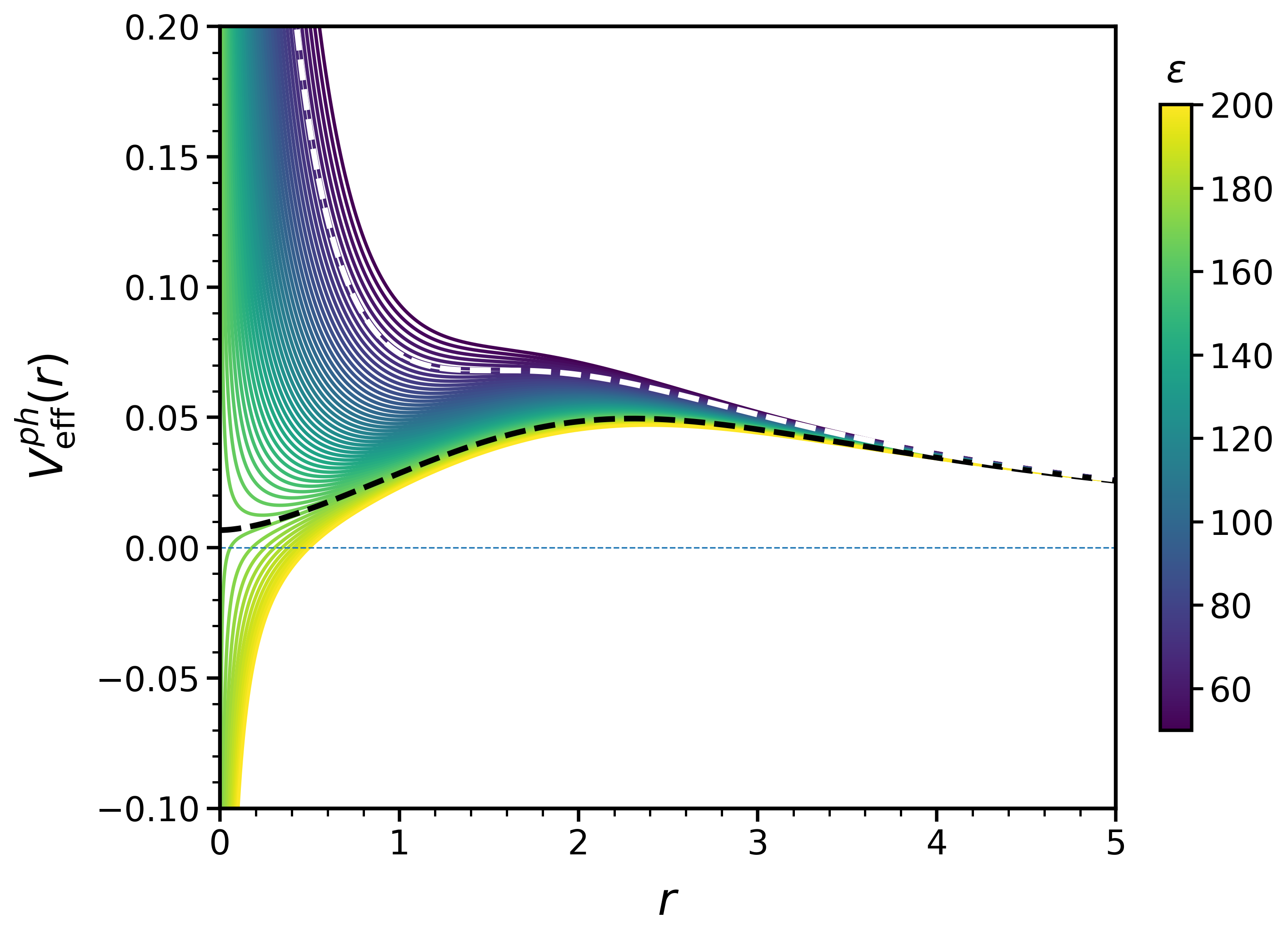}
		\end{subfigure}
		\caption{The photon effective potential $V_{\rm eff}^{\rm ph}(r)$ for various values of $\epsilon$ and $Q$, with $M=1.0$, $\kappa=8\pi$, and $\Lambda=0$: (left) $Q=0.31$ and (right) $Q=0.35$. The dashed white curve (inflection point) corresponds to the critical boundary separating configurations with no extrema from those exhibiting both a local minimum and a local maximum. The red solid curve represents a point on the extremal horizon boundary, while the dashed black curve marks a point on the non-critical boundary (zero-radius orbit). The latter separates configurations with only a local maximum from those exhibiting both a local minimum and a local maximum.}
		\label{fig:Veff_ph}
	\end{figure}

	To confirm the stability properties of the photon circular orbits explicitly, we have plotted the effective radial potential $V_{\rm eff}^{\rm ph}(r)$ from Eq.~\eqref{eq:Veff_ph_r} in Fig.~\ref{fig:Veff_ph} for various $\epsilon$, where the left and right panels correspond to $Q=0.31$ and $Q=0.35$, respectively. We find that the outer photon sphere is a local maximum of the effective potential, while the inner one is a local minimum, confirming that the inner photon orbit is stable and the outer one is unstable against radial perturbations. 
		
	As discussed in the previous subsection, the critical photon boundary does not terminate at the bifurcation point $(Q_{\rm b}, \epsilon_{\rm b})$, in contrast to the background null circular orbits, and the horizonless region supporting photon orbits spans the entire parameter space. As shown in the left panel of Fig.~\ref{fig:Veff_ph}, this region is bounded by the dashed white curve on the critical boundary $Q_{\rm m}^{\rm ph}(\epsilon;M)$ and the red solid curve on the horizon extremal boundary $Q_{\rm ext}(\epsilon;M)$, while the right panel shows the horizonless configurations bounded between $Q_{\rm nc}(\epsilon;M)$ and $Q_{\rm m}^{\rm ph}(\epsilon;M)$ boundaries for $Q = 0.35 > Q_{\rm b}$. Throughout, the potential is non-negative and develops both local minima and maxima together with an infinite barrier at $r=0$, since $Q>Q_{\rm nc}$ (equivalently $C<0$), as follows from
	\begin{align}\label{eq:Veff_ph_limit}
		\lim_{r \to 0} V_{\rm eff}^{\rm ph}(r)
		\approx	\frac{L^{2}}{\sqrt{2\epsilon}|Q|}
		\left(
		- \frac{2C(\epsilon,Q;M)}{r}
		+ 1
		-\frac{\sqrt{2}\kappa |Q|}{\sqrt{\epsilon}}
		+ \cdots
		\right),
	\end{align}
	and, hence, $V_{\rm eff}^{\rm ph}(r\to0) \propto 1/r \to \infty$. This barrier is softer than its gravitational counterpart, which diverges as $1/r^{3}$ in the same limit. The background and photon sectors share the same qualitative features: a curvature singularity at the origin, an infinite effective barrier, and a metastable trapping well at finite radii whose outer edge is set by a finite potential barrier. As in the background sector, the singularity is not removed --- the Kretschmann scalar diverges at $r=0$~\cite{Soleng1995}. Non-radial geodesics are blocked by the infinite barrier and cannot reach $r=0$, whereas radial geodesics meet no such barrier and terminate at the singularity.

	Moreover, on the non-critical boundary ($C=0$, or $Q = Q_{\rm nc} = \gamma M^{2/3} \epsilon^{1/6}$) --- marked by dashed black curves in the figure --- both inner photon orbits and inner horizons shrink to the origin at \( r=0 \). In this case, as we have seen, the background potential diverges as $1/r^{2}$ --- or as $-\ln(r)$ at the bifurcation point $\epsilon = \epsilon_{\rm b}$ --- whereas the photon potential remains finite. To see this, we substitute \( C=0 \) into the above equation and obtain:
	\begin{align}\label{eq:Veff_ph_limit_2}
		\lim_{r \to 0} V_{\rm eff}^{\rm ph}(r)
		\approx
		\frac{L^{2}}{\sqrt{2}\gamma M^{2/3}\epsilon^{2/3}}
		\left(
		1
		-\frac{\sqrt{2}\kappa \gamma M^{2/3}}{\epsilon^{1/3}}
		+ \cdots
		\right).
	\end{align}
	The LNED corrections therefore regularize the singularity of the effective potential at the origin. For $\epsilon < \epsilon_{\rm b}$, the outer horizon has a finite radius, supporting black hole solutions. For $\epsilon \ge \epsilon_{\rm b}$, however, the configuration is either a black point or a horizonless one, and in both cases it supports a finite-radius unstable photon sphere, whereas the corresponding background outer circular orbit shrinks to $r=0$. The two sectors differ in where their barriers sit. In the background sector, the potential develops an infinite barrier at $r=0$, so that the background acts as a perfect repeller. In the photon sector, by contrast, the barrier is finite and located at a finite positive radius, which is precisely what allows the unstable photon sphere to survive at finite radius. In this regime the potential is non-negative, but its minimum sits at $r=0$, where the curvature singularity lies. Consequently, no stable orbits exist at any finite radius and photons cannot be trapped, so the photon-sector configuration behaves as a partial reflector.
	
	The curvature singularity at $r=0$ is thus not regularized in either sector. Unlike regular black holes such as Bardeen or Hayward, where a de Sitter core replaces the singularity~\cite{AyonBeato1998,Hayward2006}, the present LNED solution remains singular. Whether a different Lagrangian, or quantum-gravity effects, could smooth it out, or whether such configurations can form dynamically~\cite{Vertogradov2025}, remains an open question. This mirrors the extremal Reissner--Nordstr\"om case, where the horizon shrinks to zero and the solution is known to be unstable~\cite{Aretakis2011}. We therefore treat these horizonless and black point configurations as theoretical laboratories for the nonlinear optical geometry rather than as realistic alternatives to black holes.

	\subsection{Critical Impact Parameter}\label{subsec:critical_impact}
	
	The photon effective potential analyzed in Subsec.~\ref{subsec:Veff_ph} determines not only the existence and stability of circular photon orbits but also the characteristic scale of critical photon scattering. For the outer unstable photon sphere, this scale is the critical impact parameter
	\begin{equation}\label{eq:critical_impact_ph}
		b_{\rm c}^{\rm ph}=
		\frac{\left(r_{\rm ph}^4+2\epsilon Q^2\right)^{1/4}}
		{\sqrt{f(r_{\rm ph})}},
	\end{equation}
	which follows from the first circular-orbit condition in the effective optical geometry. The physically relevant value is the one evaluated at the outer unstable photon sphere, since this orbit separates effective null trajectories that escape to infinity from those that are captured or deflected toward the central region. For comparison, the corresponding background circular-null orbit is characterized by
	\begin{equation}\label{eq:critical_impact_bg}
		b_{\rm c}^{\rm bg}=\frac{r_{\rm c}}{\sqrt{f(r_{\rm c})}}.
	\end{equation}
	The critical impact parameter is relevant to three distinct classes of configurations: black holes, for which it determines the shadow scale; black points, for which it remains finite even as the background circular-null orbit collapses; and horizonless configurations, for which it sets a critical optical scale without an event horizon.
	
\begin{figure}[t!]
		\centering
		\begin{subfigure}{0.49\textwidth}
			\centering
			\includegraphics[width=\linewidth]{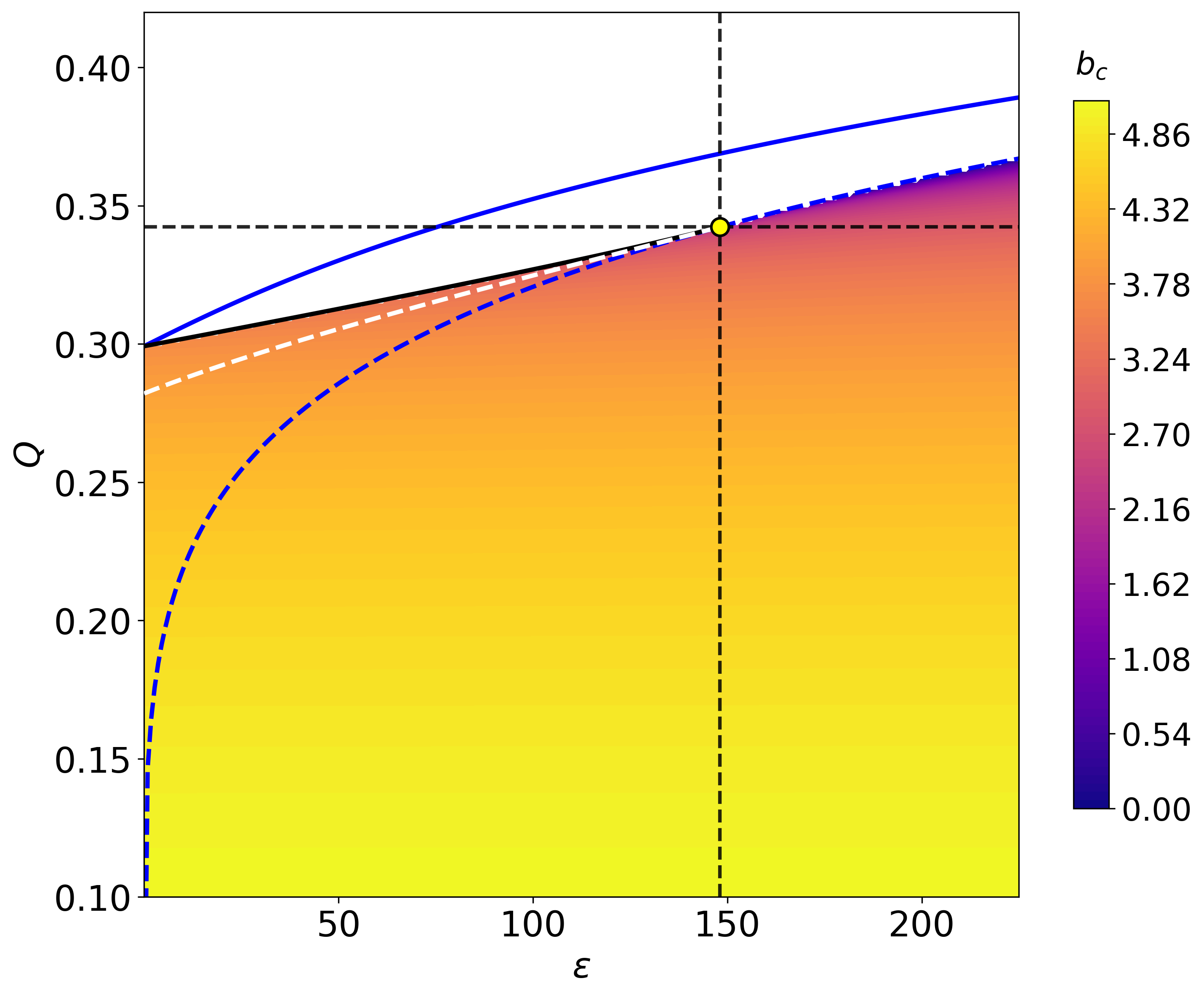}
			\caption{\footnotesize{Background null circular impact parameter}}
		\end{subfigure}
		\hfill
		\begin{subfigure}{0.49\textwidth}
			\centering
			\includegraphics[width=\linewidth]{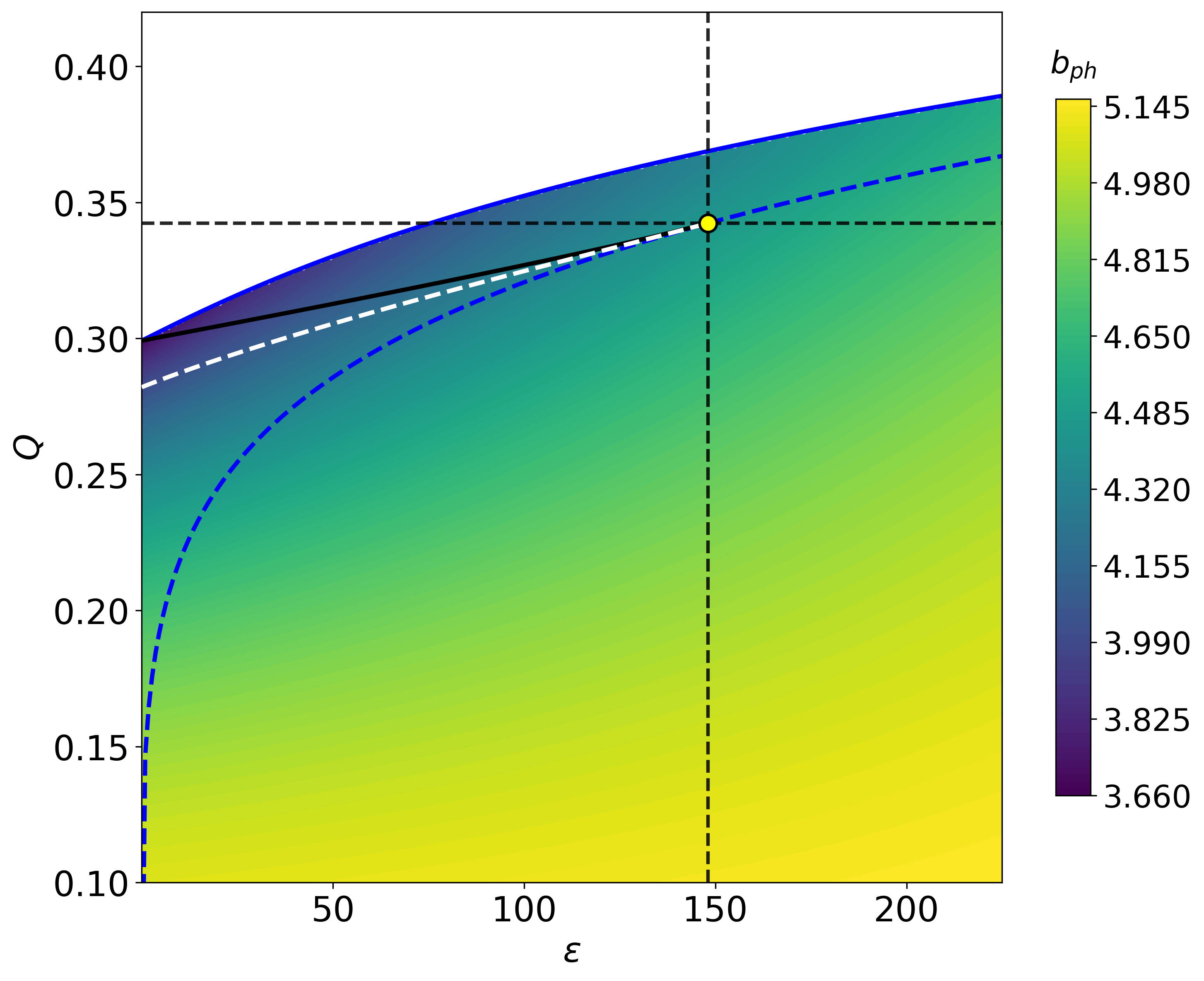}
			\caption{\footnotesize{Effective photon-sphere impact parameter}}
		\end{subfigure}
		\caption{Numerical critical impact parameters associated with the background circular-null orbit and the effective photon sphere mapped onto the $(Q,\epsilon)$ parameter space for $M=1.0$, $\kappa=8\pi$, and $\Lambda=0$. The left panel shows the impact parameter obtained from the background circular-null orbit, while the right panel shows the physically relevant impact parameter associated with the outer unstable photon sphere. The solid black curve denotes the critical boundary of the background circular-null structure, while the solid and dashed blue curves denote the numerical critical and analytical non-critical boundaries of the effective photon-sphere structure, respectively. The dashed white curve shows the extremal horizon boundary. The yellow circle marks the horizon/background zero-radius bifurcation point $(Q_{\rm b},\epsilon_{\rm b})$. It is shown for reference and is not an endpoint of the effective photon-sphere merger curve.}
		\label{fig:crit_imp_par}
	\end{figure}

	Figure~\ref{fig:crit_imp_par} shows the numerical values of these two quantities. Over the parameter range displayed, $b_{\rm c}^{\rm ph}$ is generally larger than $b_{\rm c}^{\rm bg}$. The enhancement is not due to $\Delta(r)>r^2$ alone, since the two trajectories are evaluated at different radii, and it therefore reflects both the modified optical angular sector and the displacement of the effective photon-sphere radius from the background circular-null radius. In the Maxwell limit $\Delta(r)\to r^2$ and $r_{\rm ph}\to r_{\rm c}$, so the two impact parameters coincide and reduce to the Reissner--Nordstr\"om value. The two impact parameters thus translate the phase structures of Subsecs.~\ref{subsec:Veff_bg} and~\ref{subsec:Veff_ph} into potentially observable scales.
	
	For black-hole configurations with an event horizon, $b_{\rm c}^{\rm ph}$ provides the critical scale associated with photon capture and shadow formation under appropriate observer conditions. For the black points, $b_{\rm c}^{\rm ph}$ remains finite even though the background circular-null orbit has collapsed to $r=0$, so that a finite shadow persists where a background analysis would predict its disappearance. In horizonless configurations the same critical impact parameter sets the shadow scale, but since no event horizon is present, the dark region corresponds to the photon-capture zone of the effective geometry rather than to the shadow of a horizon. This interpretation applies to all horizonless configurations considered here, irrespective of whether the effective potential develops an infinite barrier at $r=0$ or remains finite there. The coexistence of stable and unstable photon spheres found here fits into the broader class of compact-object geometries in which nonlinear electrodynamics or other strong-field modifications generate multiple circular photon orbits~\cite{Guo2023}, and the effective-geometry framework parallels the shadow analysis of Tang et al.~\cite{Tang2023}, although the present work focuses on the static phase structure rather than on rotating shadow images.

	These results suggest several observational directions. For black-hole configurations, the difference between $b_{\rm c}^{\rm ph}$ and $b_{\rm c}^{\rm bg}$ shifts the critical capture scale and hence the shadow boundary inferred from a given mass and charge. Since the two impact parameters are controlled by different phase boundaries, a measurement of the shadow size alone would not uniquely determine the underlying null-orbit structure unless combined with independent information about the charge and the nonlinear coupling. For the black points, the situation is sharpest: a finite critical impact parameter persists in a regime where a background analysis would predict its disappearance entirely. In horizonless configurations, the same scale would manifest as a critical scattering scale rather than as a sharp shadow edge, and the inner stable photon sphere could in principle leave imprints on the intensity distribution or on the time delay of radiation passing through the trapping region. High-frequency electromagnetic perturbations, quasinormal modes, and birefringence in the effective optical geometry are natural next steps in this direction~\cite{Toshmatov2019,Breton2021}, particularly where two effective photon spheres coexist or where the background circular-null structure has already disappeared. We note that in NED models depending on both electromagnetic invariants, vacuum birefringence can produce two distinct light rings and shadows~\cite{dePaula2026}, in contrast to the single-invariant LNED model considered here.

	\section{Summary and Conclusion}\label{sec:conclusion}
	
	We have analyzed the horizon structure, the background null circular orbits, and the effective photon orbits of static, spherically symmetric charged black holes in logarithmic nonlinear electrodynamics (LNED). Although these three structures share the same zero-radius non-critical boundary \(Q_{\rm nc}(\epsilon;M)=\gamma M^{2/3}\epsilon^{1/6}\), they possess distinct finite-radius critical boundaries. The horizon geometry is bounded by the finite-radius extremal curve \(Q_{\rm ext}(\epsilon;M)\), obtained from \(f(r_{\rm h})=f'(r_{\rm h})=0\), and the zero-radius non-extremal curve \(C(\epsilon,Q;M)=0\). For \(\Lambda=0\), these boundaries meet at the bifurcation point \(Q_{\rm b}=9M/[2\Gamma^2(1/4)]\), \(\epsilon_{\rm b}=128\pi^2Q_{\rm b}^2\), where the extremal horizon radius vanishes and the two-horizon phase terminates. For \(\epsilon<\epsilon_{\rm b}\), single-horizon, two-horizon, and horizonless configurations are possible; for \(\epsilon>\epsilon_{\rm b}\), the two-horizon phase is absent.
	
	The background null-geodesic condition \(r_{\rm c}f'(r_{\rm c})-2f(r_{\rm c})=0\) yields inner stable and outer unstable circular-null branches. Their finite-radius coalescence defines the merger curve \(Q_{\rm m}(\epsilon;M)\), which approaches the Reissner--Nordstr\"om values \(Q_{0}=3M/(2\sqrt{\kappa})\), \(R_{0}=3M/2\) as \(\epsilon\to0\). The perturbative expansion of \(Q_{\rm m}\) about the Maxwell limit remains accurate to sub-percent level even at the bifurcation point, while the hybrid expression for \(r_{\rm m}\) captures the rapid suppression of the merger radius near the endpoint. The background circular-null structure terminates at the same bifurcation point \((Q_{\rm b},\epsilon_{\rm b})\) as the horizon geometry.
	
	The physical photon structure is qualitatively different because electromagnetic waves in LNED propagate along null geodesics of an effective optical geometry. After removing an irrelevant conformal factor, the photon dynamics is governed by \(\Delta(r)=\sqrt{r^4+2\epsilon Q^2}\) and \(V_{\rm eff}^{\rm ph}=L^2f/\Delta\), instead of the background potential \(L^2f/r^2\). The effective geometry supports inner stable and outer unstable photon-orbit branches whose finite-radius coalescence defines the critical curve \(Q_{\rm m}^{\rm ph}(\epsilon;M)\). In the Maxwell limit, \(\Delta\to r^2\), and \(Q_{\rm m}^{\rm ph}\) approaches the same Reissner--Nordstr\"om merger configuration as \(Q_{\rm m}\).
	
	The central distinction is that \(Q_{\rm m}^{\rm ph}\) does not pass through the horizon/background bifurcation point \((Q_{\rm b},\epsilon_{\rm b})\). For \(\epsilon<\epsilon_{\rm b}\), both the background and effective photon structures extend into the horizonless region, with the effective photon structure reaching farther. For \(\epsilon>\epsilon_{\rm b}\), the background circular-null structure ceases to extend into the horizonless region, whereas the effective photon-orbit structure persists for all \(\epsilon\). In particular, for \(\epsilon>\epsilon_{\rm b}\) and \(Q_{\rm nc}<Q<Q_{\rm m}^{\rm ph}\), two effective photon circular orbits coexist despite the absence of any finite positive background circular-null orbit. The disappearance of a background circular-null orbit therefore does not imply the disappearance of the physical photon structure in LNED.
	
	The effective potentials confirm this picture. On the non-critical boundary \(Q=Q_{\rm nc}\), the background potential develops an infinite barrier at \(r=0\) for \(\epsilon\ge\epsilon_{\rm b}\), with no finite-radius extrema, whereas the photon potential is regularized and retains finite-radius extrema. In the horizonless region between the non-critical and merger curves, both geometries develop an infinite barrier at the curvature singularity together with a minimum and a maximum at finite radius, making the circular photon orbits metastable; whether long-lived trapped modes exist requires a dedicated perturbation analysis. The zero-radius limit of these configurations is the black points. The bifurcation point itself is the Soleng-type black point, where \(r_{\rm h}=0\) is a genuine horizon solution on \(Q_{\rm nc}\), while the ordinary black points are reached from the black-hole side, \(Q<Q_{\rm nc}\) with \(\epsilon>\epsilon_{\rm b}\), as \(r_{\rm h}\to0\). With the exponential quantum correction included, the entropy approaches the finite value \(S_{\eta}(r_{\rm h}\to0)=\eta\), even though the curvature singularity persists. The temperature vanishes at the bifurcation point and diverges for the ordinary black points above \(\epsilon_{\rm b}\), except when \(\eta\delta=1\), in which case the extremal temperature diverges logarithmically instead of vanishing. In the small-mass limit the ordinary black points recover the Schwarzschild temperature, entropy, and heat capacity, whereas the Soleng-type black points do not.
	
	The critical impact parameters quantify the distinction between the two orbit structures. For the outer unstable effective photon sphere, \(b_{\rm c}^{\rm ph}=\sqrt{\Delta(r_{\rm ph})/f(r_{\rm ph})}\), whereas for the background circular-null orbit, \(b_{\rm c}^{\rm bg}=r_{\rm c}/\sqrt{f(r_{\rm c})}\). Their difference reflects both the displacement of the orbit radius and the modification of the angular sector by the effective optical geometry. For black-hole configurations, \(b_{\rm c}^{\rm ph}\) sets the scale for photon capture and shadow formation; for the black points, \(b_{\rm c}^{\rm ph}\) remains finite even though the background circular-null orbit has collapsed to \(r=0\), so that a finite shadow persists where a background analysis would predict its disappearance. In horizonless configurations, it should be interpreted as a critical optical impact parameter rather than automatically as a black-hole shadow radius.
	
	These results connect naturally with several earlier studies. The background circular-null and effective photon-orbit structures found here are consistent with the bound-orbit and stable photon configurations reported by Habibina et al.~\cite{Habibina2021} and with the multiple photon spheres discussed by Guo et al.~\cite{Guo2023}. The effective-geometry treatment parallels the slowly rotating shadow analysis of Tang et al.~\cite{Tang2023}, although the present work isolates the static circular-photon structure and its phase boundaries rather than the rotating shadow images. The horizon phase structure is in line with the Hendi-type logarithmic black-hole family~\cite{Hendi2013} and with dyonic LNED solutions~\cite{Kruglov2019,Luo2024}, while the thermodynamic and stability properties of related LNED black holes have been explored by Aram et al.~\cite{Brzo2026}. The persistence of circular-orbit structures in horizonless configurations echoes similar findings in regular black-hole and NED spacetimes~\cite{Stuchlik2015,Chiba2017,Rayimbaev2020,Toshmatov2026}. In particular, the horizonless configurations found here---featuring an infinite barrier at the curvature singularity, metastable wells at finite radius, and coexisting stable and unstable effective photon orbits---share qualitative features with other horizonless compact objects supported by nonlinear electrodynamics, although the specific mechanism differs. The absence of a de Sitter core distinguishes the present solution from the Bardeen and Hayward regular black holes~\cite{AyonBeato1998,Hayward2006}, and the instability of the zero-radius extremal configuration mirrors the known instability of extremal Reissner--Nordstr"om black holes~\cite{Aretakis2011,Angelopoulos2024}. In contrast to the regular spacetimes constructed in~\cite{Wang2026}, where boundedness of the Kretschmann scalar is imposed by construction, the present LNED solution retains a curvature singularity at $r=0$ while still supporting photon spheres and horizonless configurations, showing that regularity is not a prerequisite for the existence of a photon sphere or a critical impact parameter.
	
	Several directions for future work follow directly from these results. First, the coexistence of stable inner and unstable outer effective photon circular orbits motivates a detailed study of high-frequency electromagnetic perturbations, quasinormal modes, and possible long-lived trapped modes in the effective optical geometry~\cite{Toshmatov2019,Breton2021}. The metastable trapping regions identified here provide a natural setting for such analyses, particularly in the horizonless regime where the background circular-null structure has disappeared. Second, the phase-structure analysis can be extended to dyonic and magnetically charged LNED configurations~\cite{Kruglov2019,Luo2024}, Born--Infeld black holes~\cite{Demianski1986,Dey2004,Cai2004}, and other logarithmic NED families~\cite{Hendi2013}, allowing universal and model-dependent boundary relations to be identified. Third, the critical impact parameters computed here provide a basis for comparison with shadow observations, and extending the present static analysis to rotating and asymptotically non-flat LNED configurations would allow direct contact with the slowly rotating results of Tang et al.~\cite{Tang2023}. Fourth, the dynamical formation and stability of the horizonless configurations, together with the fate of the black points and the curvature singularity, remain open questions~\cite{Vertogradov2025, Battista2026b}. The present treatment includes the exponential quantum correction to the entropy without backreaction on the geometry, so whether backreaction, a different Lagrangian, or a full quantum-gravity treatment could alter the singularity or the stability of these zero-radius configurations deserves further investigation~\cite{Aretakis2011,Angelopoulos2024}.
	
	The distinction between background null geodesics and effective photon trajectories established here provides a general framework for interpreting strong-field electromagnetic propagation in nonlinear electrodynamics, and it suggests that the phase structure of the effective optical geometry---rather than that of the background spacetime alone---should guide the interpretation of photon trapping, critical scattering, and shadow-related observables in such theories.

	\section*{Acknowledgment}

	\printbibliography
	
\end{document}